\documentclass{emulateapj}
\usepackage{graphicx}
\newcommand{\ha}        {\mbox{H$\alpha$}}
\newcommand{\hi}        {\mbox{\rm \ion{H}{1}}}
\newcommand{\hii}       {\mbox{\rm \ion{H}{2}}}
\shortauthors{Sheets et al.}
\shorttitle{Dusty OB Stars in SMC. I}
\begin{document}
\title{Dusty OB Stars in the Small Magellanic Cloud - I: Optical Spectroscopy
       Reveals Predominantly Main-Sequence OB Stars}
\author{Holly A. Sheets$^1$,
        Alberto D. Bolatto$^1$,
        Jacco Th. van Loon$^2$,
        Karin Sandstrom$^3$,
        Joshua D. Simon$^4$,
        Joana M. Oliveira$^2$,
        and Rodolfo H. Barb\'{a}$^5$}
\affil{
$^1$Department of Astronomy, University of Maryland, College Park, MD
    20742-2421, USA; bolatto@astro.umd.edu\\
$^2$Lennard-Jones Laboratories, Keele University, ST5 5BG, UK\\
$^3$Max-Planck Institut f\"ur Astronomie, K\"onigstuhl 17, D-69117 Heidelberg,
    Germany\\
$^4$Observatories of the Carnegie Institution for Science, 813 Santa Barbara
    Street, Pasadena, CA 91101, USA\\
$^5$Departamento de F\'{\i}sica, Universidad de La Serena, Cisternas 1200
    Norte, La Serena, Chile}
\begin{abstract}
We present the results of optical spectroscopic follow-up of 125
candidate main sequence OB stars in the Small Magellanic Cloud (SMC)
that were originally identified in the S$^3$MC infrared imaging survey as
showing an excess of emission at 24~\micron\ indicative of warm dust,
such as that associated with a transitional or debris disks. We use these
long-slit spectra to investigate the origin of the 24 $\mu$m emission
and the nature of these stars. A possible explanation for the observed
24~\micron\ excess, that these are emission line stars with dusty
excretion disks, is disproven for the majority of our sources. We find
that 88 of these objects are normal stars without line emission, with
spectral types mostly ranging from late-O to early-B; luminosity
classes from the literature for a sub-set of our sample indicate that
most are main-sequence stars. We further identify 17 emission-line
stars, 7 possible emission-line stars, and 5 other objects with
forbidden-line emission in our sample. We discover a new O6\,Iaf star;
it exhibits strong \ion{He}{2} 4686~\AA\ emission but relatively weak
\ion{N}{3} 4640~\AA\ emission that we attribute to the lower nitrogen
abundance in the SMC. Two other objects are identified with planetary
nebulae, one with a young stellar object, and two with X-ray
binaries. To shed additional light on the nature of the observed
24~\micron\ excess we use optical and infrared photometry to estimate
the dust properties of the objects with normal O and B star spectra
and compare these properties to those of a sample of hot spots in the
Galactic interstellar medium (ISM). We find that the dust properties
of the dusty OB star sample resemble the properties of the Galactic
sample of hot spots.  Some may be runaway systems with bow-shocks
resulting from a large velocity difference between star and ISM.  We
further investigate the nature of these dusty OB stars  in a companion paper
presenting mid-infrared spectroscopy and additional imaging.
\end{abstract}

%============================================================================ 1
\section{Introduction}
\begin{figure}
\epsscale{0.5}
\plotone{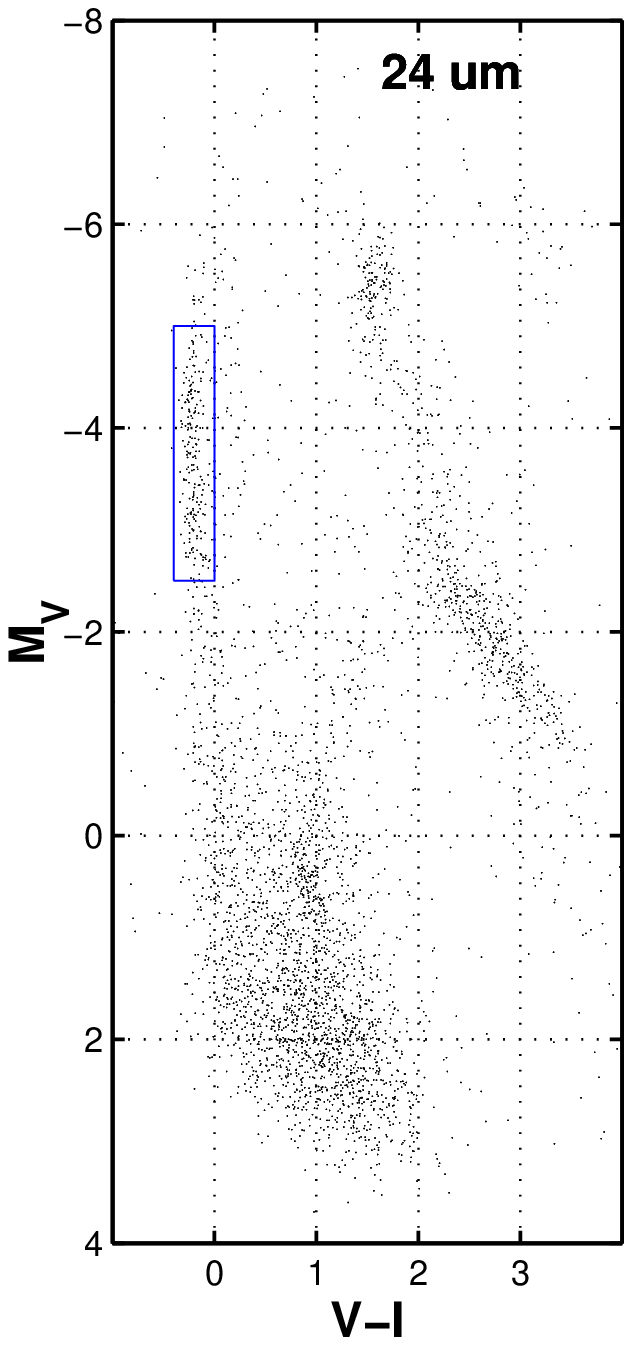}
\caption{$V, V-I$ color--magnitude diagram for all of the stars detected at
24~\micron\ in the S$^3$MC survey. The box indicates the objects with $M_V$
and $V-I$ consistent with main-sequence late-O and early-B stars ($-5 < M_V <
-2.5$ mag and $-0.4 < V-I < 0$ mag). The photospheres of such stars should not
be detectable at 24~\micron\ in the S$^3$MC survey.}
\label{fig:only24}
\end{figure}

The {\it Spitzer} Survey of the Small Magellanic Cloud (S$^3$MC)
carried out deep imaging of the main body of the SMC in all seven IRAC
and MIPS bands \citep{bolatto}. This survey identified about 400,000
compact sources in the infrared. Compilation of spectral energy
distributions and cross-correlation against optical catalogs uncovered
193 point sources with 24-\micron\ emission, but with optical and
near-infrared colors and magnitudes all consistent with O9--B3 type
main-sequence stars. Photospheres of such stars are well below the
detection limit of this survey, so an additional source of infrared
(IR) emission must be present, presumably warm dust with a temperature
close to 150~K. Figure~\ref{fig:only24} shows the $V, V-I$
color--magnitude diagram for all objects detected at 24~\micron\ in
the S$^3$MC survey, and the box represents the $V-I$ and $M_V$ cuts on
the dusty OB star sample. These sources are a puzzle, representing a
few percent of all the stars in that particular color and magnitude
range in the SMC. Dust in close proximity to an early B/late O star
could be a remnant from the accretion process, and provide important
information about the final stages of accretion and the clearing up of
the original disk in massive stars. We discuss below several competing
possibilities for the nature of these sources, previously presented by
\citet{bolatto}. The discovery of these objects has been confirmed
independently by the {\it Spitzer} SAGE-SMC survey \citep[][who also
discuss them]{sagesmc} and by \citet{akari} using AKARI data.

If the dust is associated with the star, it must exist either in an
optically thin shell surrounding the star or in a thin or transition
circumstellar disk, since no appreciable reddening of the central star
is detected beyond the line-of-sight extinction by our own Galaxy to
the SMC. These objects have $F_{IR}/F_{Bol}\sim10^{-4}$ to 10$^{-2}$,
indicating that the dust is intercepting only a small fraction of the
star's light and re-radiating it in the IR. This also supports the
thin shell or disk scenario. Since most of the dusty OB stars show
little to no excess emission at wavelengths of 8~\micron\ or shorter,
a large central gap would be needed to explain the absence of
significant emission from hotter dust located close to the star.

A clump of interstellar dust, heated by a nearby star, however, could
also reproduce these characteristics. This is observed, for example,
in the Pleiades \citep{arny77,wb93,hs01,sloan04}. Vega-like stars,
Herbig Ae/Be stars, and classical Be stars all have disks that are
known to contain dust to varying degrees, and so we consider them, as
well as cirrus hot spots in the ISM, as the possible causes for the
excess emission seen around these O and B stars in the S$^3$MC survey.
 
Vega-like stars host debris disks, which are the remnants of planet
formation.  They are optically thin dust disks, mostly stripped of
their primordial gas and continually replenished by collisions between
planetesimals. The prototypical examples are Vega, Fomalhaut, and
$\beta$~Pictoris, all of which have disks that extend 100 to 1000 au
from the star with L$_{IR}$/L$_{Bol}$ $\sim10^{-5}$ to 10$^{-3}$
\citep[][and references therein]{back}. The central region, a few au
in extent, of these disks has been cleared of dust, just as in our
solar system, and the mass of emitting dust is $\sim$ 10$^{-3}$ to
10$^{-2}$ M$_{\earth}$ at temperatures between 50 and 125 K. The wider
sample of debris disks shows an upper limit on F$_{IR}$/F$_{Bol}$ of
10$^{-2}$ \citep{chen06}, temperatures up to 300~K, and dust masses as
large as an Earth mass \citep{krivov}. Note however that the known
debris-disk hosts are observed, overwhelmingly, around much less
massive stars with later spectral types. A gas-free disk around
a very luminous star would be quickly cleared by radiation pressure,
unless the dust grains are very large.
% NOTE: find the broader range of temps, for asteroid belt analogs and Kuiper
% belt.
% Maybe note that these have been resolved, and give parameters for the larger
% sample of unresolved?

Classical Be stars are rapidly rotating main-sequence stars with an
excretion disk of gaseous material \citep{port}. The disk produces IR
excess in the form of free--free emission with approximately
$S_{\nu}\propto\nu^{0.6}$ in the near- and mid-IR \citep{waters0};
this by itself would not explain the sudden rise observed in the
spectral energy distribution (SED) of the dusty O and B stars at
24~\micron. \citet{mirbjor}, however, found evidence for
circumstellar, thermal dust emission in some Be stars in the far-IR,
which they attribute to remnant dust from prior evolutionary
processes.

Herbig Ae/Be stars, on the other hand, are pre-main sequence stars that are
thought to be the more massive analogs to T~Tauri stars. They are surrounded
by gaseous accretion disks containing dust with temperatures as high as
1500~K, creating the IR excess \citep[see, e.g.][]{hill,waters}. Herbig Ae/Be
stars typically show significant excess at $\lambda\geq 1$~\micron\ from this
hot dust. The onset of the emission at longer wavelengths indicates that the
central portion of the disk nearest the star has been cleared and 
is free of dust. Disks with such a central hole are transition disks
\citep[e.g.,][]{cieza}. It is possible that more massive, early-B and O
stars also go through this stage, although modeling suggests that this
may be an exceedingly short phase as disks are rapidly cleared
\citep{alexander}. Observational evidence for circumstellar disks
around massive B and O stars remains hard to come by \citep{cesaroni}.

The Galactic cirrus, comprised of wispy or filamentary patches of dust
throughout the disk of the Galaxy, has a typical temperature of $\sim18$~K
\citep{planck24}, but heating of the cirrus by embedded stars produces small
regions of warmer dust, or hot spots, near the star, reaching temperatures of
$\sim70$~K \citep{vanmc}. It is difficult to distinguish debris disks from hot
spots in the cirrus without resolved images of either the dust emission in the
IR, or the scattered light in the optical. The observed SEDs can be
interpreted with either scenario, as illustrated by \citet{mart} and
\citet{su06}; Mart\'{\i}nez-Galarza et al.\ model some of the same objects in
Su et al.\ as ISM heated by stars passing through a cloud, while Su et al.\
model them as debris disks. Cirrus hot spots are also sometimes the cause of
the far-IR emission detected around Be stars \citep{mirbjor}.

Understanding the source of the excess emission in these objects is
valuable because the scenarios above represent different phases in the
evolution of massive stars. Detecting circumstellar material around
SMC stars is interesting, because it offers the opportunity to study
accretion or excretion processes in a low metallicity environment, and
particularly around massive stars with an unextincted
line-of-sight. Detecting debris disks in the SMC would be especially
exciting, as they would offer an unprecedented opportunity to study
planet formation in another galaxy. Such early-type stars are
under-represented in Galactic studies.

While the photometry suggested the stellar counterparts of the IR
emission to be early-type main-sequence stars, this has yet to be
confirmed through spectroscopic observation. The vast majority of the
193 objects in our sample do not have spectral types reported in the
literature, although a small subsample has fiber optics spectroscopy
\citep{evans}.

\citet{sagesmc} compiled a catalog of IR counterparts to 
massive stars spectroscopically confirmed in the literature. They
identify 44 objects that are similar (OB stars with 24 $\mu$m
emission) but not necessarily identical to our sample. They 
constitute $\sim2$\% of their sample of late-O and early-B stellar
types, a fraction slightly lower but otherwise similar to that
reported by \citet{bolatto}. The 18 spectra
\citeauthor{sagesmc} have in hand show nebular line emission, unlike --- as we shall
see --- the majority of the sources in this sample. Note that the
spectra used by \citeauthor{sagesmc} are obtained using fibers
\citep{evans,evans06}, which makes sky subtraction in complex regions
problematic. Long-slit spectroscopy is preferred in regions of diffuse
sky emission such as found around massive stars. Clearly, further
spectroscopic study is warranted. \citet{sagesmc} conclude that most
of their sources are not found within very young regions, and suggest
their IR emission is likely dust associated with cirrus or a nearby
molecular cloud, rather than disks. \citet{bolatto}, however, noted
that the majority of the sources identified in their sample are in the
vicinity of active star-forming regions although not in their cores,
where confusion with diffuse 24 $\mu$m emission would make their
identification extremely difficult.

%due to the heavy extinction
%encountered toward the Galactic star-forming regions in which they are
%located. 
%They also offer an opportunity to study such objects at the
%low metallicity of the SMC. 
We obtained optical long-slit spectroscopy for 125 of the 193 objects
from the S$^3$MC survey in order to determine how many of them are
emission-line (e.g., classical Be and Herbig Ae/Be) stars and to
obtain spectral types to quantify the stellar radiation field
illuminating the dust. In section 2 of this paper, we discuss the
photometric and spectroscopic observations, while in section 3 we
classify the stars, and in section 4 we estimate the properties of the
dust around the non-emission line objects. In section 5, we discuss
how well the various scenarios explain our objects and include a
comparison of the dust properties to those estimated from the sample
of Galactic cirrus hot spots of
\citet{gaust}. In a companion paper \citep[][hereafter Paper
II]{adams}, we discuss further infrared spectroscopic and photometric
information, and comparisons to large Galactic samples based on 
{\em Wide-field Infrared Survey Explorer} (WISE)
data.

%============================================================================ 2
\section{Observations}
%-------------------------------------------------------------------------- 2.1
\subsection{Spectroscopy}

The optical spectra were obtained at the 3.5-m New Technology Telescope at the
European Southern Observatory at La Silla, Chile in 2007 September (programme
079.C-0485; PI J.Th.van Loon), using the EMMI instrument \citep{emmi} in RILD
mode in the red arm with grism \#2 and a $1\rlap{.}\arcsec 0$ slit. This
set-up provided wavelength coverage from $\sim385$ to 870~nm with
$\lambda/\Delta\lambda=570$. Biases, dome flats, and HeAr arc lamp frames were
acquired for calibration. The science integrations were split into three
exposures, offset along the slit, in order to correct for cosmic rays and
fringing effects at wavelengths beyond 750~nm). A filter to block second-order
contamination longward of 800~nm was originally planned, but it was dropped
because of unwanted reduction of signal in the blue region of the spectrum.

The spectra were wavelength-calibrated with HeAr lamp exposures, using the
IRAF\footnote{IRAF is distributed by the National Optical Astronomy
Observatories.} {\sc identify}, {\sc reidentify}, and {\sc transform} tasks in
the {\sc noao} spectroscopic reduction package. The 1D spectra were then
extracted using the {\sc apsum} task. The extraction was at times complicated
by the background emission lines from \hii\ regions within which our targets
were located. To remove this background, the aperture and background regions
were set interactively, looking first at the profile of the star along the
slit, away from any emission lines. If any [\ion{O}{3}] 5007~\AA\ emission
remained in the extracted spectrum, the extraction was re-done looking instead
at the profile along the slit at the [\ion{O}{3}] line to set the aperture and
background regions. The extracted spectra were summed for each object to
increase the number of counts and to average out fringing effects and an
electronic interference pattern that could not be removed with the bias or
dark frames due to frame-to-frame variations. Lastly, the spectra were
normalized to the continuum in the region between 400 and 680~nm by fitting
and dividing by an order 20 polynomial.

%------------------------------------------------------------------------- 2.2
\subsection{Photometry}

The photometry used to create the SEDs for each source is a combination of
{\it B}, {\it V}, and {\it I} data from the OGLE~II survey \citep{ogle} and
the Magellanic Clouds Photometric Survey \citep{zar}, {\it J}, {\it H}, and
{\it K$_{s}$} from the 2MASS survey \citep{mass}, and {\it Spitzer} Space
Telescope \citep{spitzer} data from both the original S$^{3}$MC
\citep{bolatto} survey and the combined S$^{3}$MC/SAGE-SMC catalogs
\citep{gordon11}, making use of the IRAC \citep{irac} and MIPS \citep{mips}
instruments. In the SAGE-SMC processing, combined S$^{3}$MC/SAGE-SMC
images were created by stacking data from both surveys before
performing the photometry. For the MIPS combined catalog, on the other
hand, photometry was performed on the three different observing epochs
(one for S$^{3}$MC, and two for SAGE-SMC) separately, after uniformly
reprocessing the data. We searched the combined catalogs for
coordinate matches to the dusty OB stars within 2\arcsec. The IRAC and
2MASS data points in the original S$^{3}$MC photometry were replaced
with the points in the combined catalog if they existed. If more than
one match was found, the closest match was used. The 24~\micron\
points in the original S$^{3}$MC catalog were replaced with the
weighted average of the three epochs in the combined catalog for the
the closest positional match. Points were replaced as long as there
was a match within 2\arcsec\ in at least one of the epochs. The
70~\micron\ data points, where they exist, must be treated with
caution. The diffuse emission in the SMC is considerably brighter at
70~\micron\ than at 24~\micron, and the difference in beam size
between the two channels is substantial. The beam size at 24~\micron\
is 6\arcsec, corresponding to 2~pc at the SMC distance of 61.1~kpc
\citep{kw06}, while the beam size at 70~\micron\ is 18\arcsec,
corresponding to 5~pc. Thus there is a good chance of source confusion
between the two channels, and we have not searched the combined
S$^{3}$MC/SAGE-SMC catalog for 70~\micron\ matches. The photometry
data are given in Tables~\ref{tab:phottab} and~\ref{tab:phottab2}.

%=========================================================================== 3
\section{Spectral Classification}

\begin{figure}
\epsscale{1.0}
\plotone{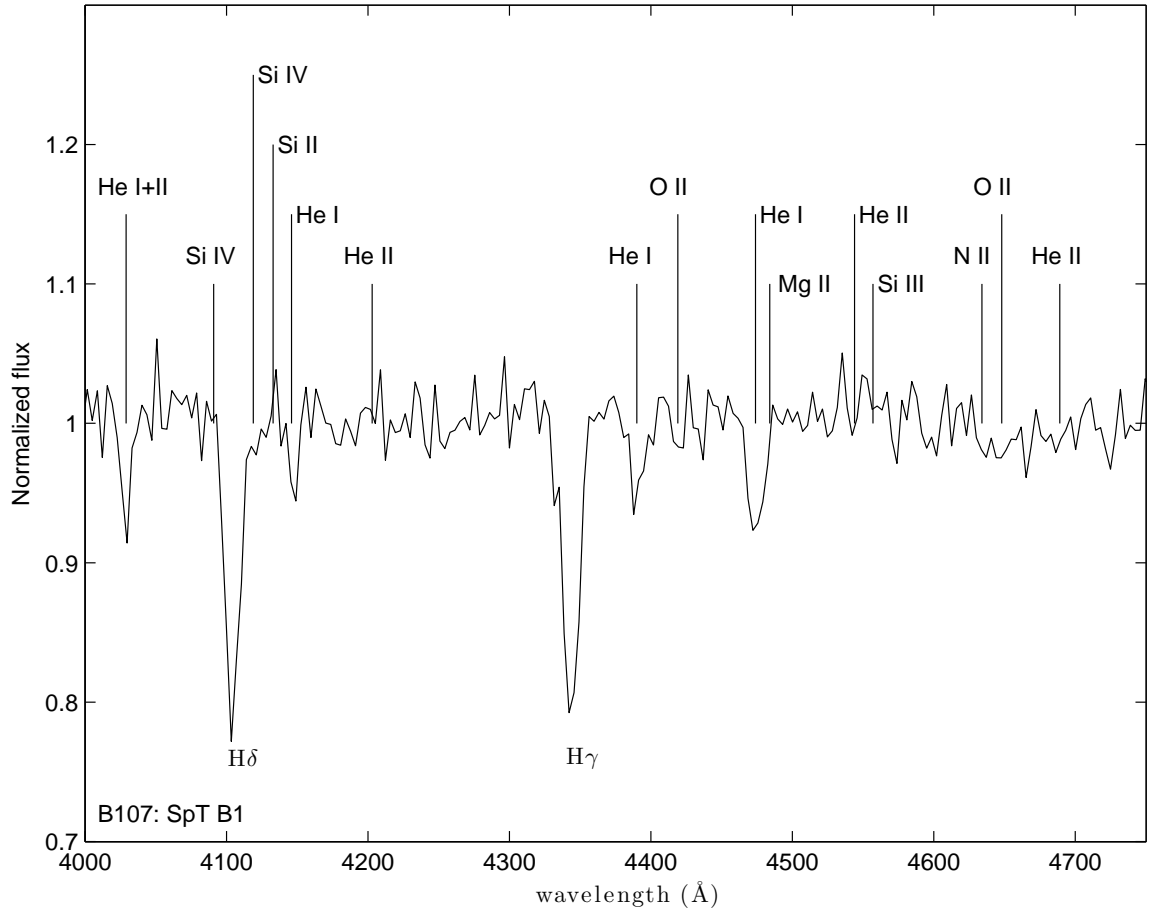}
\caption{Example optical spectrum of one of the SMC dusty OB stars, B\,107. To demonstrate the
classification system, the important lines are marked. The lack of
\ion{He}{2} lines indicates that it is a B star, and the lack of
\ion{Si}{3} lines suggests it is not a B2/B3 star. The \ion{Mg}{2}
line, however, is too weak for a star later than B3, so we conclude
that the star is a B1, with the
\ion{Si}{4} lines blended with H$\delta$.}
\label{fig:samp}
\end{figure}

We use our long-slit spectroscopy to classify the stars. For the
classification, we employ the scheme described in \citet{evans}, which
compares the strengths of a number of lines in the 4000--5000 \AA\
range.
% The strong presence of He II 4686 \AA\ indicates the star is an O star, and
% the strength of the He II 4200 \AA\ line relative to several other lines
% indicates whether it is an O9, O8.5, O7, or O6 star.
We use the equivalent widths of the lines, guided by visual inspection of the
spectra, to compare the line strengths, but our low resolution leaves some of
the important lines blended with H$\delta$ (\ion{Si}{4} 4088~\AA\ and
\ion{Si}{4} 4116~\AA) or with each other (e.g., \ion{He}{1} 4471~\AA\ and
\ion{Mg}{2} 4481~\AA). Note also that the original classification scheme was designed
for supergiants, and so the Si lines for main-sequence stars may be
too weak to detect in some cases. The classification scheme is
summarized in Table~\ref{tab:class}. A sample spectrum of object
B\,107 is shown in Figure~\ref{fig:samp}. Since B\,107 shows no
\ion{He}{2} absorption features, it is not an O star. It shows no
evidence of \ion{Si}{3} 4553~\AA, so it is not a B2 or B3, but the
\ion{Mg}{2} 4481~\AA\ appears too weak for a spectral type later than
B3. Thus we conclude that most likely the \ion{Si}{4} lines are
blended with H$\delta$ and that the star is a B1.

\begin{figure}
\epsscale{1.0}
\plotone{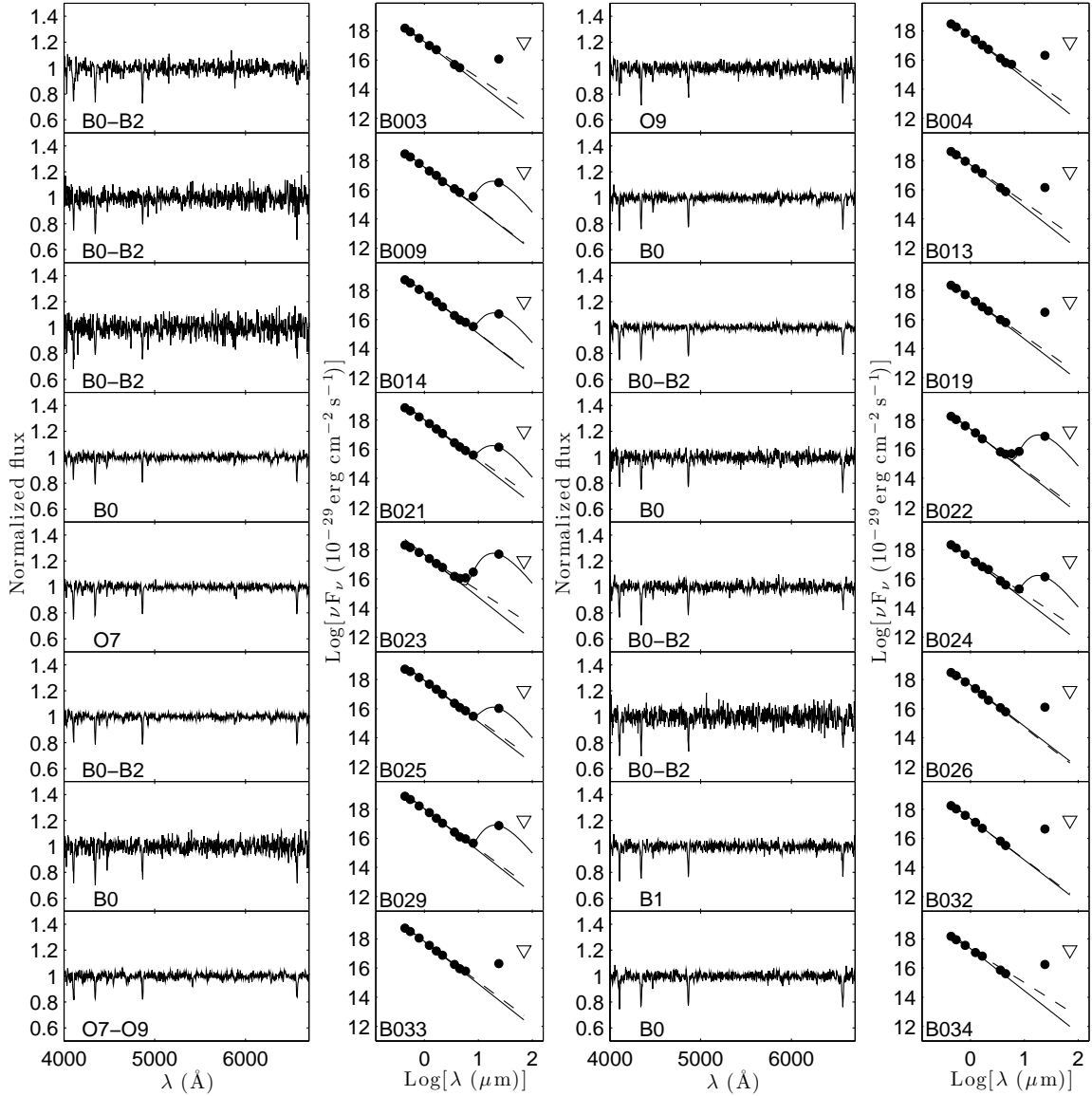}
\caption{Spectra and SEDs of the objects with spectra of normal late-O and
early-B stars. The solid line represents the fit of Equation 1 to the data
using the slope from the spectral type, scaled by the $J$ flux, or $I$ if $J$
was unavailable. Stars with a type O7--O9 use a slope of $-2.91$, which is the
slope for all stars in that range, while stars with type B0--B2 use a slope
corresponding to B1. The dashed line is a fit of Equation 1 to the $I$--$K$
points, when available, without a fixed slope. Also plotted as a solid line is
the total emission from the dust plus the photosphere, in cases where the
8/24~\micron\ color temperature was determined. The open triangle indicates
the flux limit at 70~\micron, and, for the few objects with a 70~\micron\
detection, the total dust plus photosphere emission for the 24/70~\micron\
dust temperature is also plotted. Important lines are those marked in
Figure~\ref{fig:samp}, as well as \ha\ at 6563~\AA, H$\beta$ at 4861~\AA,
and \ion{He}{1} 5876~\AA. The [\ion{O}{3}] 5007~\AA\ line was used to gauge
the effectiveness of the subtraction of the background \hii\ region. The
$B$--$H$ fluxes have been corrected for the foreground extinction by the Milky
Way.}
\label{fig:norm1}
\end{figure}

\begin{figure}
\epsscale{1.0}
\plotone{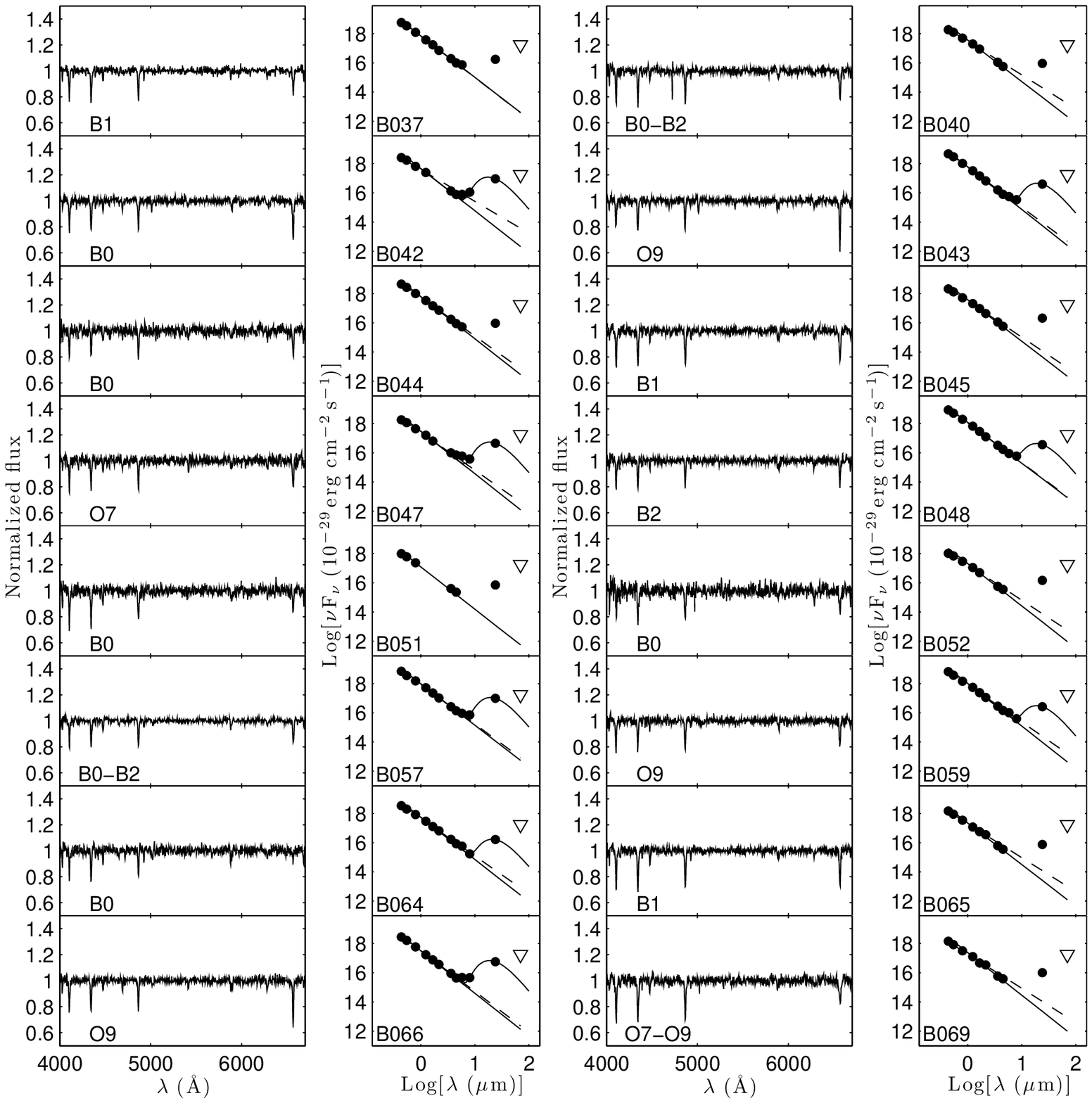}
\caption{Same as Figure~\ref{fig:norm1}.}
\label{fig:norm2}
\end{figure}

\begin{figure}
\epsscale{1.0}
\plotone{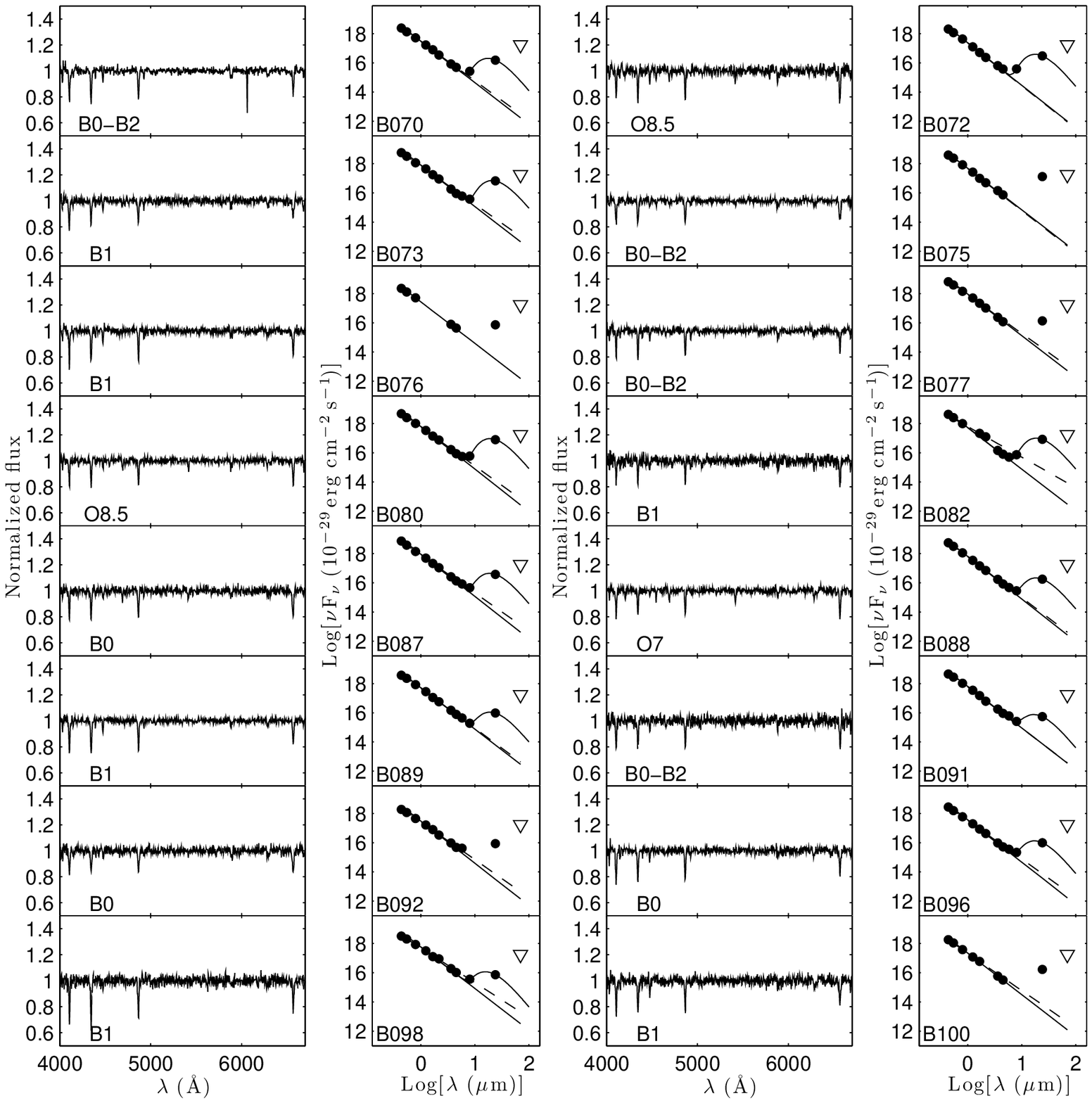}
\caption{Same as Figure~\ref{fig:norm1}.}
\label{fig:norm3}
\end{figure}

\begin{figure}
\epsscale{1.0}
\plotone{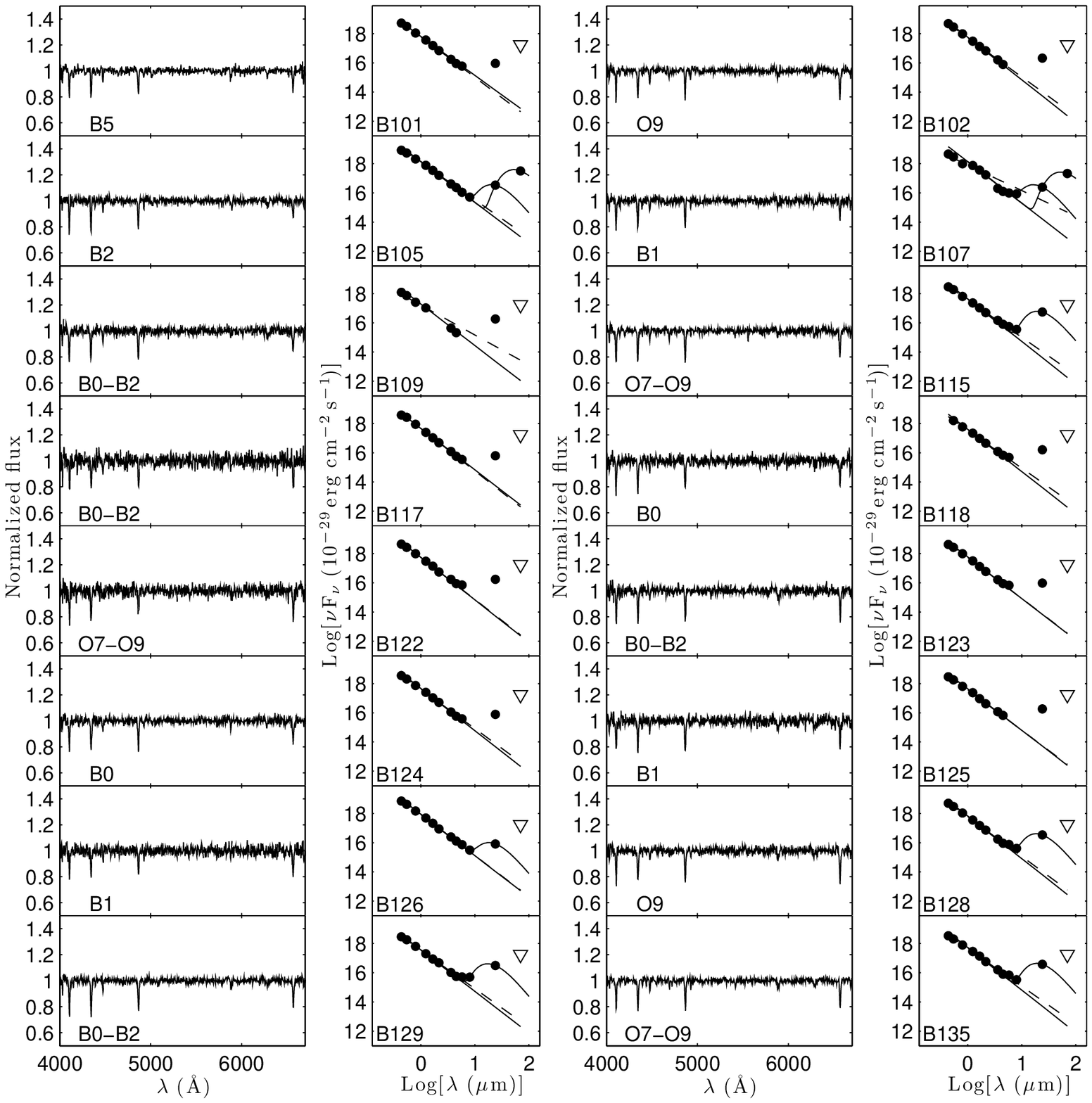}
\caption{Same as Figure~\ref{fig:norm1}.}
\label{fig:norm4}
\end{figure}

\begin{figure}
\epsscale{1.0}
\plotone{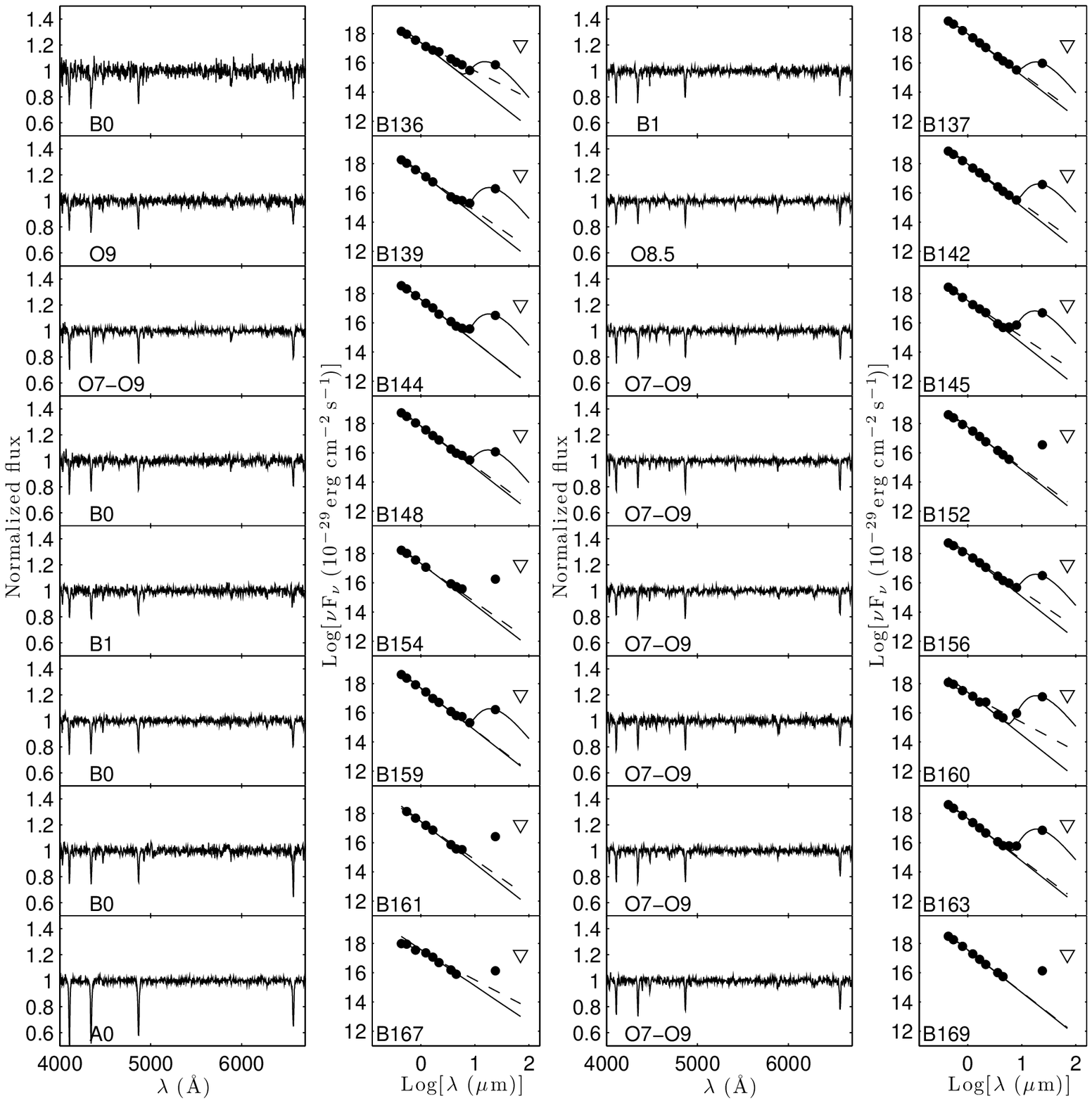}
\caption{Same as Figure~\ref{fig:norm1}.}
\label{fig:norm5}
\end{figure}

\begin{figure}
\epsscale{1.0}
\plotone{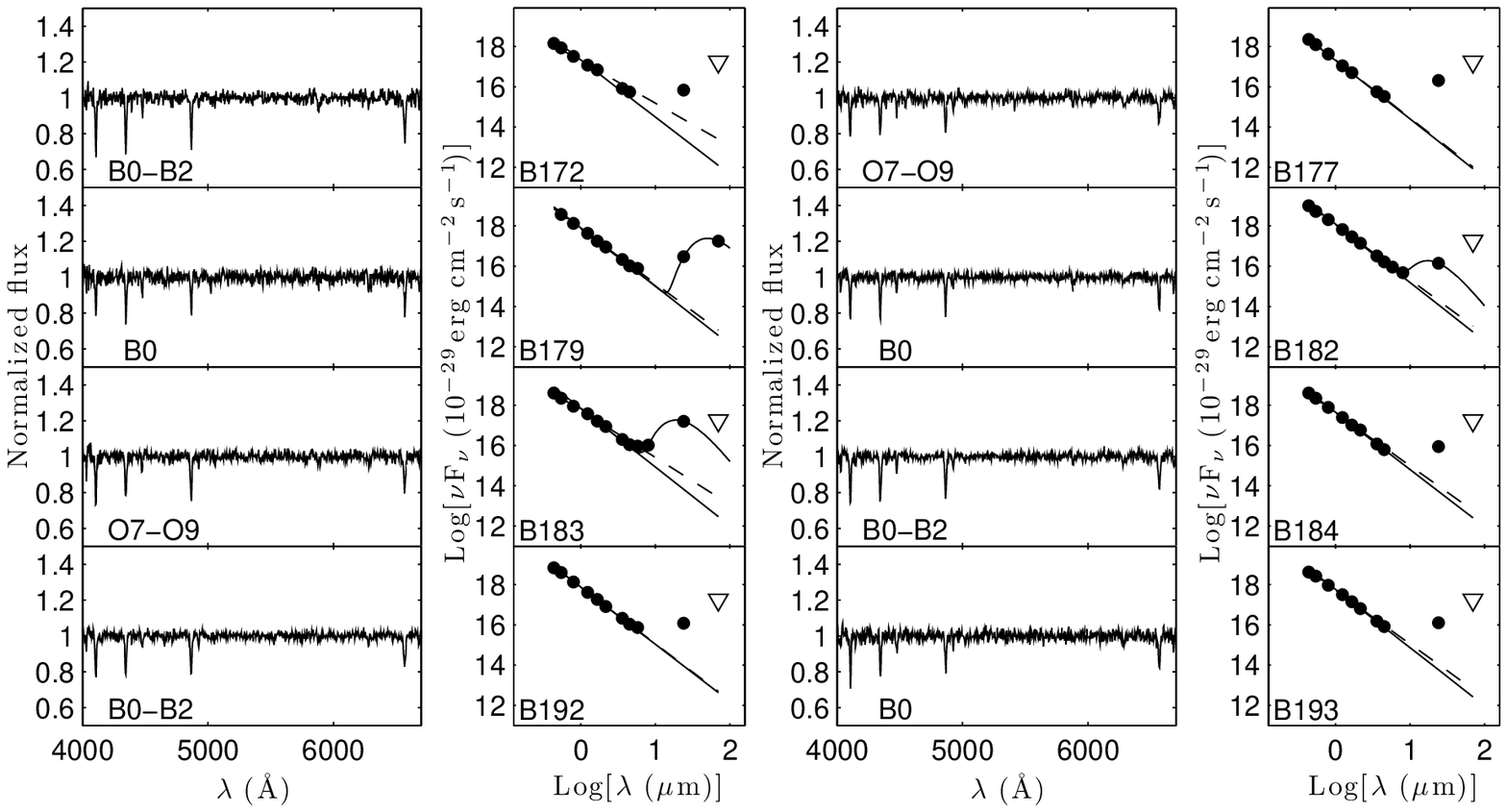}
\caption{Same as Figure~\ref{fig:norm1}.}
\label{fig:norm6}
\end{figure}

Of the 125 objects for which we obtained spectra, 87 appear to be
normal main-sequence stars and one (B\,167) is discussed below as a
likely supergiant.  Of those 87, 53 are classified via the above
scheme and assigned a subtype.  The remaining 34 stars were difficult
to place precisely within the scheme, but they have been marked as
either O7--O9, if they show \ion{He}{2} absorption, or B0--B2 if they
lack the \ion{He}{2} lines. B\,167 is unusual, however, in that it
shows no significant \ion{He}{1} absorption but has very strong Balmer
absorption. It appears to be an A star; given its luminosity it is
probably a supergiant. The spectra and SEDs of these 88 objects are
shown in Figures~\ref{fig:norm1} through~\ref{fig:norm6}, while the
assigned spectral type is given in Table~\ref{tab:spec} as well as in
the figures.
% NOTE: B167 is unusual in that it has strong Balmer absorption but no He I
% absorption, or other lines really, until out around 8500 \AA, where it looks
% like some strong paschen lines at about 8551, 8600, 8671, 8753, 8867, 9018,
% and 9237 \AA - those wavelengths do not account for the redshift of the SMC
% which is roughly 3 \AA in the blue. It's located in a complicated part of
% the SMC. Rodolfo called it an A star.

The remaining 37 of the 125 objects, which are not considered in the dust
analysis in the next section, fall into one of four groups: 1) stars that show
Balmer emission and sometimes \ion{He}{1}/\ion{}{2} emission (16 stars plus B\,094), 2)
stars with unusual \ha\ profiles (7 stars, including B\,035 which we especifically discuss
in the following section), 3) objects that show strong
forbidden line emission (6 objects), and 4) spectra that are difficult to
extract (8 stars).
 
\begin{figure}
\epsscale{1.0}
\plotone{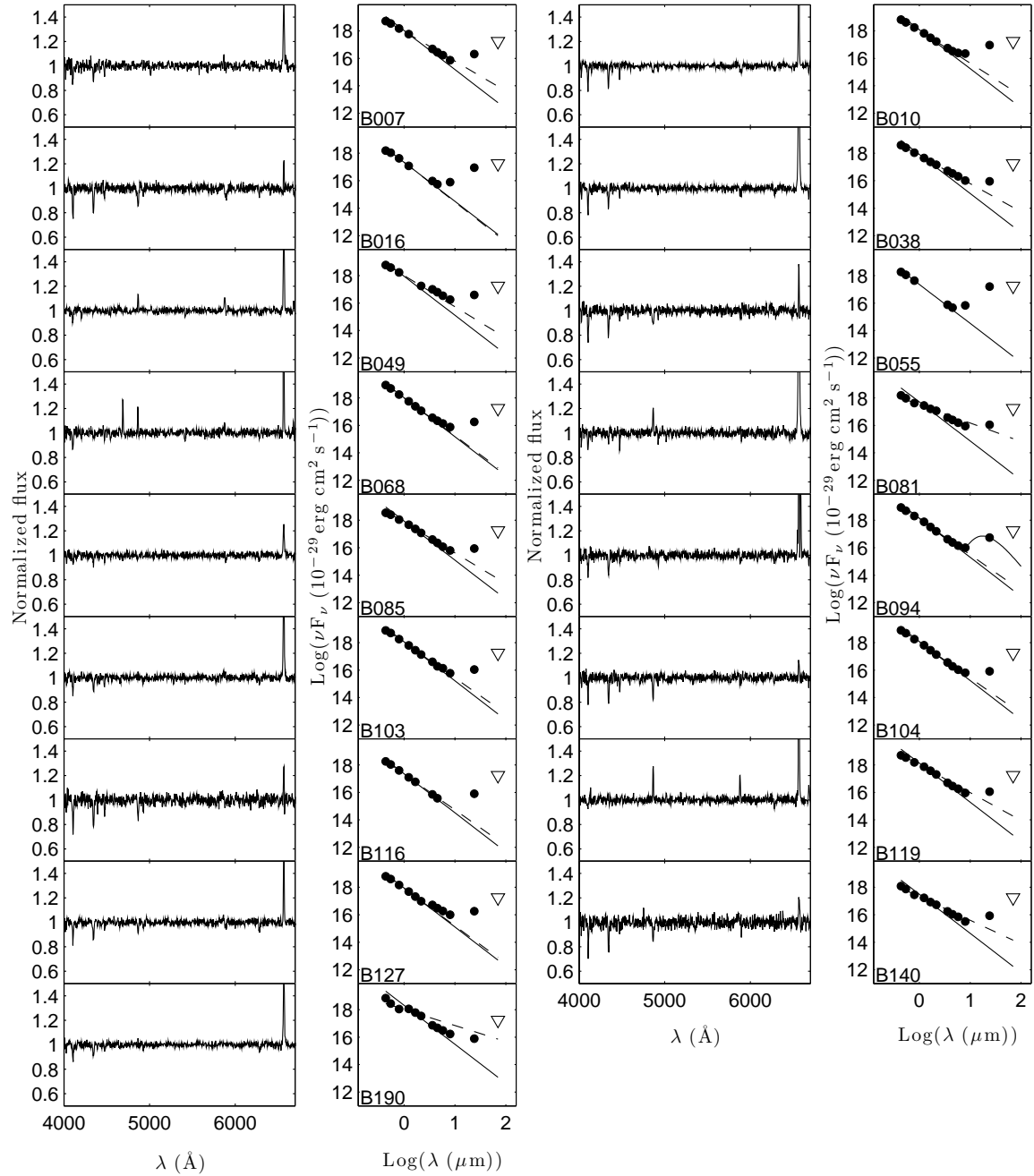}
\caption{Spectra and SEDs, as in Figure~\ref{fig:norm1}, of the objects that
we identify as Be stars based on the Balmer emission with no [\ion{O}{3}]
5007~\AA\ emission, a line whose absence we take to indicate that background
subtraction is complete. The slopes of the solid lines in the SEDs correspond
to a spectral type of B1, because the potential for \ion{He}{1}/{\sc ii}
emission makes Be stars difficult to classify. B\,094 is unusual in that it
shows [\ion{N}{2}] emission, but no [\ion{O}{3}]. We place it in the Be-star
group because the \ha\ emission, with the rest of the Balmer series in
absorption, more closely resembles the Be stars than the group with strong
Balmer and [\ion{O}{3}] emission.}
\label{fig:be}
\end{figure}

\begin{figure}
\epsscale{1.0}
\plotone{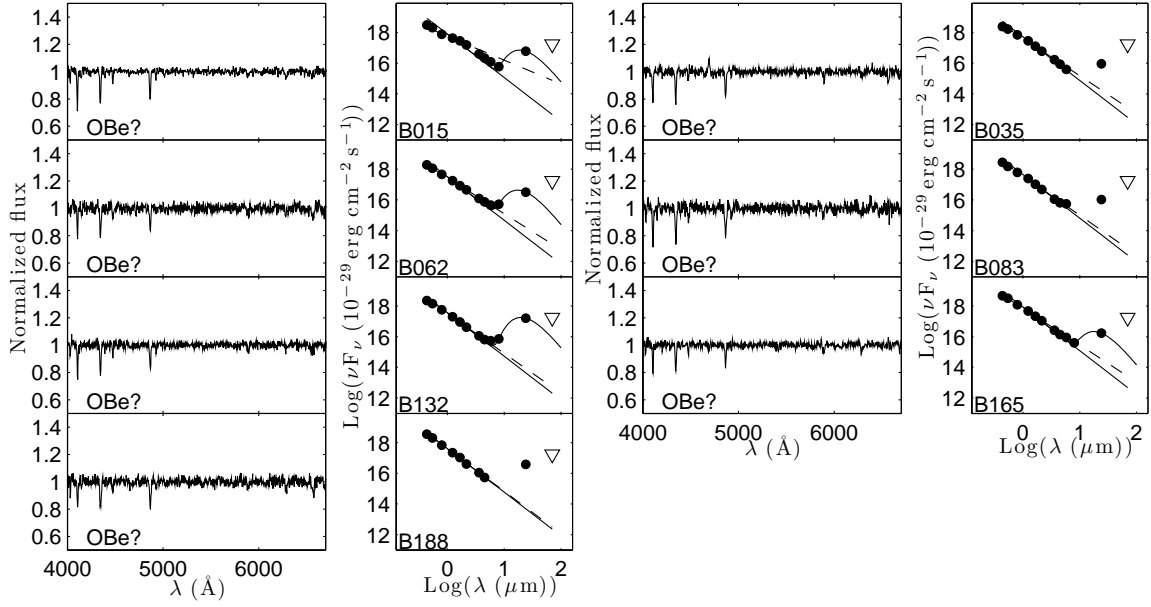}
\caption{Spectra and SEDs, as in Figure~\ref{fig:norm1}, of the objects that
show unusually little absorption at \ha\ and are potentially Be stars whose
photospheric \ha\ absorption has not been filled in completely by the emission
of the circumstellar matter. The slopes of the solid lines in the SEDs are set
as in Figure~\ref{fig:be}.}
\label{fig:flat}
\end{figure}

The first group of 17 stars shows the Balmer and \ion{He}{1}/{\sc ii} emission
lines characteristic of Be stars (Figure~\ref{fig:be}), while the second group
of 7 stars shows only weak \ha\ absorption (Figure~\ref{fig:flat}). B\,094
stands out, as it is the only object among the 17 stars
that shows [\ion{N}{2}] $\lambda\lambda$~6548,6583 emission. We include it
here, rather than with the strong forbidden-line emission objects, because of
the lack of [\ion{O}{3}] emission and the weakness of the Balmer emission
compared to the other objects in that group. The unusually weak Balmer
absorption in the second group of stars may indicate that these objects are Be
stars as well. \citet{silaj} modeled the \ha\ profiles of Be stars and showed
that, for many combinations of base-disk density, density distribution, and
viewing angle, the \ha\ emission from the circumstellar matter does not cancel
out the absorption line of the star completely. This effect could explain the
unusual \ha\ profiles. The definite Be stars are noted with a spectral type of
OBe in Table~\ref{tab:spec}, while stars with possible weak emission are noted
as ``OBe?'' in the table.

\begin{figure}
\epsscale{1.0}
\plotone{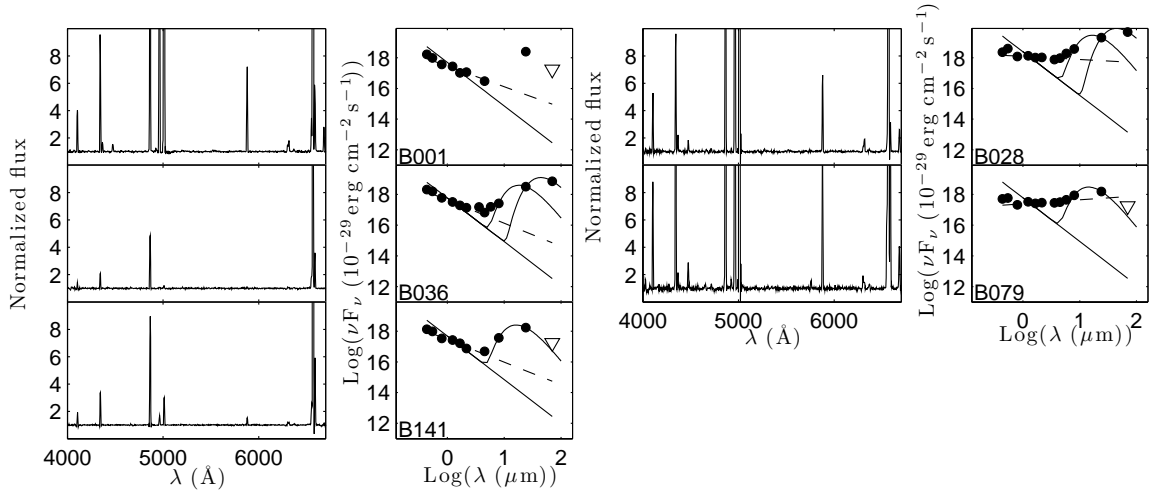}
\caption{Spectra and SEDs, as in Figure~\ref{fig:norm1}, of the forbidden-line
emission objects. The slopes of the solid lines in the SEDs correspond to a
spectral type of B1. B\,001, B\,028, B\,079, and B\,141 are believed to be
compact \hii\ regions, given the strength of the forbidden emission, while
B\,036 has been identified as a young stellar object \citep{oliveira}.}
\label{fig:chii}
\end{figure}

The third group is comprised of 6 objects (5 shown in
Figure~\ref{fig:chii}, plus the already discussed B\,094), and it
includes B\,001, B\,028, and B\,079, all of which show strong Balmer
and [\ion{O}{3}] $\lambda\lambda$~4959,5007 emission, along with
weaker [\ion{N}{2}] $\lambda\lambda$~6548,6583 and \ion{He}{2}
$\lambda\lambda$~5876,4921, and 4471~\AA\ emission. These spectra are
typical of compact \hii\ regions. B\,141 is dominated by Balmer
emission, with weak [\ion{N}{2}] and very weak
\ion{He}{1} emission. B\,141 also shows moderate [\ion{O}{3}] emission, but
unlike B\,001, B\,028, and B\,079, the forbidden-line emission is weak
relative to H$\beta$. B\,036 is now known to be a young stellar object (YSO)
\citep{oliveira}; it shows strong Balmer emission, with weak [\ion{N}{2}]
emission, very weak [\ion{O}{3}] emission, and little to no \ion{He}{1}
emission above the continuum level. The SEDs of these five sources are much
flatter than those of the non-emission-line objects, possibly due to the
predominance of the free--free emission over any stellar photospheric
emission. These objects are marked in Table~\ref{tab:spec} as Em in the
spectral type column.

\begin{figure}
\epsscale{1.0}
\plotone{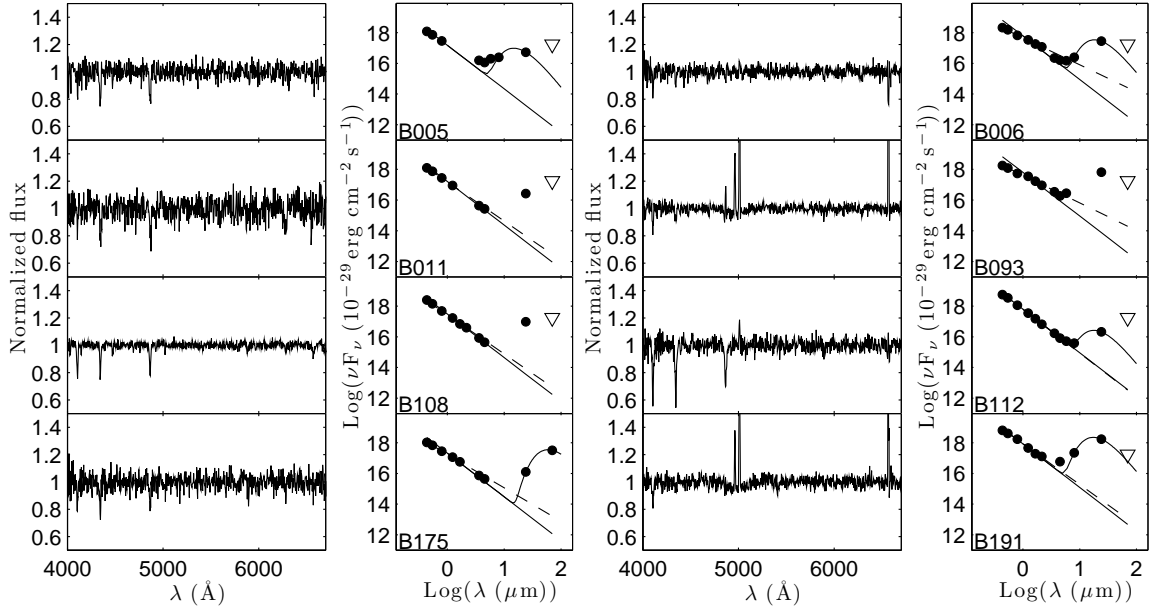}
\caption{Spectra and SEDs of the objects for which background subtraction was
difficult. B\,005, B\,011, B\,112, and B\,175 all suffer from low
signal-to-noise due to clouds, and, except for B\,011, they all have other
objects nearby on the slit. B\,006, B\,093, B\,108, and B\,191 are embedded in
a complex region of \ha\ emission, making it difficult to determine the
appropriate background level to subtract from the stellar spectrum. The slopes
of the solid lines in the SEDs again correspond to a type of B1, given the
lack of a classification.}
\label{fig:diff}
\end{figure}

The fourth group, shown in Figure~\ref{fig:diff}, includes B\,005, B\,011,
B\,112, and B\,175, all of which have low signal-to-noise spectra and, except
for B\,011, also have a star too nearby to set the background well. Thus it is
unclear whether the \ha\ absorption appears weak because of noise, background
contamination, or weak Be star emission.
% Put B005 back in with the normal stars? Rodolfo has it as an early B.
The other 4 objects in this group are B\,006, B\,093, B\,108, and B\,191,
whose complicated background in \ha\ makes it difficult to disentangle
emission from the surrounding nebula from possible Be star emission. These
objects are given no spectral type in Table~\ref{tab:spec}.

%-------------------------------------------------------------------------- 3.1
\subsection{A newly discovered O6\,Iaf star}

\begin{figure}
\epsscale{1.0}
\plotone{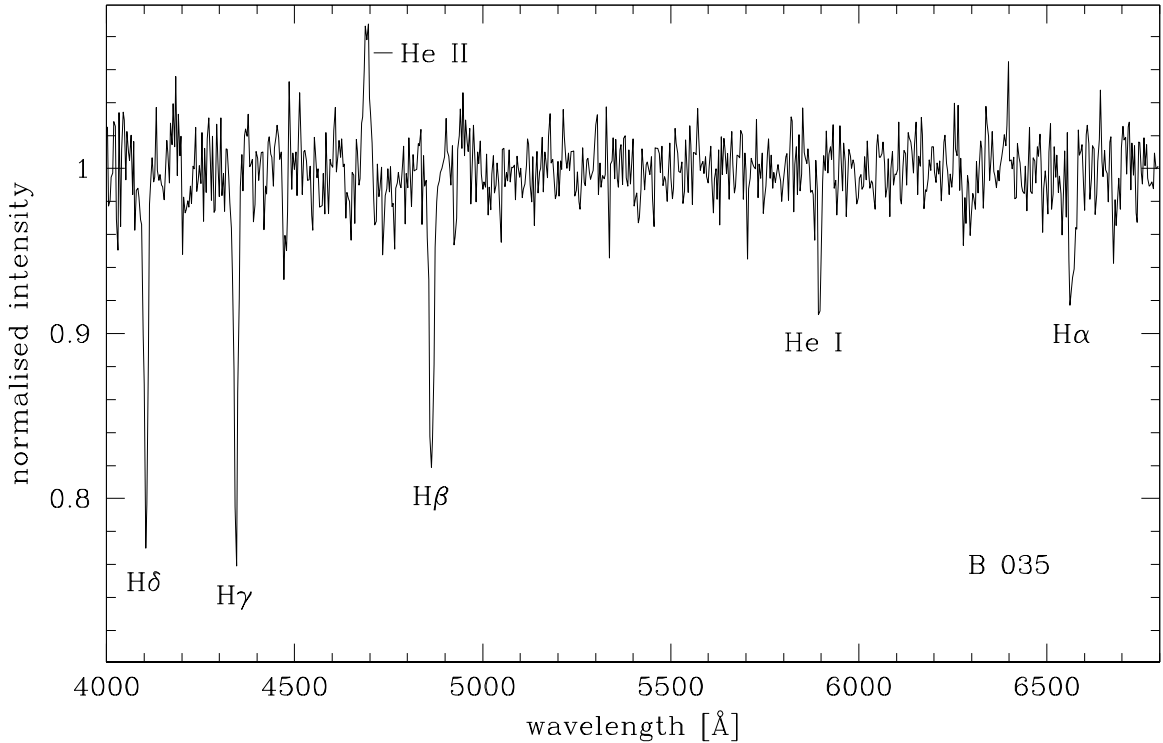}
\caption{Spectrum of B\,035, which exhibits \ion{He}{2} 4686~\AA\ emission
indicative of an Ia supergiant luminosity class. The relative strength of the
\ion{He}{1} and \ion{He}{2} lines suggests a spectral type around O6; the
\ion{N}{3} 4640~\AA\ complex is weak, possibly due to the low metallicity of
the SMC.}
\label{fig:b035}
\end{figure}

In the ``OBe?'' group, B\,035 is the only star that shows \ion{He}{2} 4686~\AA\
emission. The spectrum is displayed in more detail in Figure~\ref{fig:b035}.
The appearance of the \ion{He}{2} 4686~\AA\ emission indicates a very high
luminosity, class Ia \citep{sota}. This is commonly accompanied by emission
of the \ion{N}{3} 4640~\AA\ complex, but this is weak in the spectrum of
B\,035. We suggest that this is a metallicity effect due to the low nitrogen
abundance in the SMC \citep{vanloon10}. It does appear to be present, hence we
tentatively affix the ``f'' suffix to the spectral type. With regard to the
temperature class, the \ion{He}{1} 4471~\AA/\ion{He}{2} 4200~\AA\ ratio
suggests a spectral type of about O6 \citep{sota}. The \ion{He}{1}+{\sc ii}
4026~\AA\ line is weaker than that of the \ion{He}{2} 4200~\AA\ line, which
would suggest a spectral type $<$ O4, but it is also weaker than that of the
\ion{He}{1} 4471~\AA\ line, which would suggest a spectral type $>$ O7 (which
does not really happen, \citealt{sota}) -- we attribute this to uncertainties
and/or possible binarity and arrive at a final spectral classification of
O6\,Iaf.
% We also note the Diffuse Interstellar Bands (DIBs) in the spectrum of
% B\,035, where those at 4725, 4762 and 4780~\AA\ are relatively strong
% compared to the usually strong 4428, 5780 and 6614~\AA\ bands
% \citep{vanloon13}.
% Spectrum seems to have changed (!) No evidence for DIBs remains... (?)

%-------------------------------------------------------------------------- 3.2
\subsection{Comparison with literature}

Accurate spectral classification based on high quality data was
carried out by \citet{evans06}, \citet{martayan07}, \citet{evans08}
and \citet{hunter08} for 23 of our 125 stars (see
Table~\ref{tab:spec}). Our own spectral classifications are in
agreement with those determinations, with only a few deviations by one
or two subclasses. The majority ($\sim14$) of those classifications
are of luminosity class IV--V, with only 6 of luminosity class II and
none more luminous. This suggests that most of the stars in our sample
are unevolved, main-sequence stars.

The sample, however, is a mixed bag of curious and seemingly very common types
of objects. The ``Em'' objects include the planetary nebulae B\,001 =
LHA\,115-N\,9 and B\,079 = LHA\,115-N\,47, the YSO B\,036 \citep{oliveira},
and the compact \ion{H}{2} region B\,028 = LHA\,115-N\,26 \citep{testor}. The
sample further includes the X-ray binaries B\,064 = SXP\,214 \citep{coe11} and
B\,085 = CXOU\,J005245.0$-$722844 \citep{antoniou09} and no less than 10 or 11
eclipsing binaries. We attribute the large number of eclipsing binaries to the
high overall fraction of close binary systems among massive stars \citep{sana12,sana13}.

%============================================================================ 4
\section{Constraining the Dust Properties}

In order to assess the viability of the various hypotheses for the
nature of the dusty OB stars and the origin of the dust, we now use the
available photometry to estimate the gross properties of the dust. The
primary quantities we can constrain are the dust temperature, dust
mass, the fraction of the star's light re-radiated by the dust
(F$_{IR}$/F$_{Bol}$), and the distance of the dust from the star. We
also calculate the grain size below which radiation pressure will
quickly remove the dust from the system in the absence of gas drag, as
well as the mass of grains of that blow-out size that would be needed
to account for the observed F$_{IR}$/F$_{Bol}$.

To determine these properties, we first find the amount of expected flux from
the stellar photosphere by fitting to the SED a power law of the form
\begin{equation}
\label{nuinu}
\ln(\nu I_\nu)=m\ln(\lambda)+b
\end{equation}
with $\nu$ in Hz and $\lambda$ in \micron, and with the slope $m$ derived from
the Kurucz stellar atmosphere model \citep{castelli} of the spectral type
determined. The intercept $b$ was determined using the $J$-band flux, or $I$
if $J$ was unavailable, to scale the models. The $J$ and $I$ fluxes were
corrected for the line-of-sight extinction due to the Milky Way using
$A_I=0.482A_V$ and $A_J=0.282A_V$ from \citet{dustk} and \citet{extum}, with
$A_V=0.12$~mag \citep{schlegel}. We selected models for $[M/H]=-0.5$, closest
among the Kurucz library to the typical abundance of young stars and gas in
the SMC of $[M/H]\sim-0.7$ \citep{pagel78}, and the suggested surface
temperatures and surface gravities for specific stellar types from
\citet{martins} and Schmidt-Kaler \citep{kaler}. All the fits used models with
$\log g=4$ for the surface gravity, and the surface temperature and slope used
for each spectral type is given in Table~\ref{tab:class}. The temperature
scale is similar -- well within 10\% -- to that of Galactic main-sequence
stars determined by \citet{nieva} and that of O-type stars in the SMC
determined by \citet{massey}; the latter found that early-O type stars in the
SMC are hotter than Galactic stars of the same spectral type, but that this
difference vanishes around B0. A slope $m=-2.91$ was used if the star was
given a spectral type of ``O7--O9'', while a slope $m=-2.84$ was used for all
other stars without a specific spectral type. From these fits, the excesses at
4.5, 8, and 24~\micron\ were calculated (Table~\ref{tab:excess}), along with
the ratio of the excess to the expected photosphere.

%-------------------------------------------------------------------------- 4.1
\subsection{Dust Temperature}

In all but six cases, marked with a footnote in Table~\ref{tab:excess}, no
emission was detected at 70~\micron, so we cannot fit a model to the 8, 24,
and 70-\micron\ excesses directly. Instead we estimate the color temperature
of the dust using the 8- and 24-\micron\ excesses, and place limits on the
color temperature using the 24~\micron\ flux and the 70~\micron\ flux limit of
40,000 $\mu$Jy, minus the expected photospheric flux. The color temperatures
are estimated assuming a modified blackbody with emissivity
$\propto\lambda^{-2}$, using the equation
\begin{equation}
\label{abrat}
\frac{f_a}{f_b}=\left(\frac{\lambda_b}{\lambda_a}\right)^5
                \frac{\exp{(hc/\lambda_b kT)}-1}{\exp{(hc/\lambda_a kT)}-1},
\end{equation}
where $a$ is the shorter wavelength and $b$ the longer. The color temperatures
calculated for the ratios of the 8- to 24-\micron\ excesses, for stars with a
specific spectral type, are given in Table~\ref{tab:ourdust}. Because the
8-\micron\ band includes a major polycyclic aromatic hydrocarbon (PAH) feature,
these color temperatures must be treated with caution. Note, however, that PAH emission
in the SMC is considerably weaker than in the Milky Way making this less of a
problem than for local samples \citep{smcpah}. The color temperatures
calculated with the 24- to 70-\micron\ ratios are very uncertain, since in
most cases we have only an upper limit on the 70-\micron\ flux, and thus are
not included in the table. These values are generally in the range of 40--60
K, but they should be considered as lower limits as a 70-\micron\ flux below
the upper limit would lead to a higher derived temperature. Mid-IR
spectroscopy and deeper 70-$\mu$m photometry to determine the dust temperature
directly are presented in Paper II.

%-------------------------------------------------------------------------- 4.2
\subsection{Dust Mass}

We are working in the optically thin limit at long wavelengths, so the
emission of the dust is given by
\begin{equation}
\label{inu}
I_\nu=\Omega \tau_\lambda B_\nu,
\end{equation}
where $I_\nu$ is the excess emission at 24~\micron, $\Omega$ is the solid angle
subtended by the source, $\tau_\lambda$ is the opacity due to the dust, and
$B_\nu$ is the emission of a blackbody at the temperature of the dust.
% \citep{dust2}.
The opacity of the dust is
\begin{equation}
\label{tau}
\tau_\lambda=\frac{M_d}{\Omega D^2}\kappa_\lambda,
\end{equation}
where $M_d$ is the mass of the dust, $\Omega D^2$ is the area covered by the
dust, and $\kappa_{\lambda}$ is the dust mass absorption coefficient. Thus the
dust mass is given by
\begin{equation}
\label{mass}
M_d=\frac{I_\nu D^2}{\kappa_\lambda B_\nu}.
\end{equation}
Here we assume
\begin{equation}
\label{kappa}
\kappa_\lambda=10\left(\frac{250}{\lambda}\right)^2,
\end{equation}
\noindent where $\kappa_\lambda$ has units of cm$^2$ g$^{-1}$ and $\lambda$ is
given in
\micron. This value is due to \citet{hildebrand83} and within 20\% of \citet{dustk}. We adopt it to be easily comparable to surveys of the Milky Way \citep[for example, the Gould Belt Survey;][]{andre}. The dust mass then is
% $$M_d=1.9\times 10^{5}
%       \frac{I_{\nu(mJy)}D^2_{pc}\lambda^2_{\mu m}}
%            {B_{\nu(erg~cm^{-2} s^{-1} Hz^{-1})}}$$ or, in Earth masses,
\begin{equation}
\label{masse}
M_d=3.9\times10^{-23}\frac{I_\nu D^2\lambda^2}{B_\nu}~{\rm M_\earth},
\end{equation}
where $I_\nu$ is given in mJy, $D$ in pc, $\lambda$ in \micron, and $B_\nu$ in
erg cm$^{-2}$ s$^{-1}$ Hz$^{-1}$ sr$^{-1}$. The dust masses calculated using temperatures the
derived from 8/24-\micron\ excess ratios are listed in Table~\ref{tab:ourdust}.

%-------------------------------------------------------------------------- 4.3
\subsection{Dust Luminosity}

To calculate the fractional luminosity F$_{IR}$/F$_{Bol}$, we first determined
the infrared flux by integrating over the source function for graybody
emission from 0.7~$\mu$m to 100~$\mu$m, rewriting Eq.~\ref{inu} above as
\begin{equation}
\label{inuint}
I_\nu=3.18\times10^{-4}\frac{M_d}{D^2}\frac{1}{\lambda^2}B_\nu,
\end{equation}
where $M_d$ is in M$_\earth$, $D$ is the distance to the source in pc,
$\lambda$ in \micron, and $I_\nu$ in erg cm$^{-2}$ s$^{-1}$
Hz$^{-1}$. We then divided by the bolometric flux, which is the
luminosity for the spectral type (Table~\ref{tab:class}), scaled to
the distance of the SMC.

%-------------------------------------------------------------------------- 4.4
\subsection{Distance of the Dust from the Star}

% need to make clearer that the blackbody distance and mass estimate from the
% blowout grain size are for comparison with the debris disk model, since the
% grains in that case are large, more like blackbodies, while the modified
% blackbody distance and mass from color temp is the ISM hotspot model -
% though that mass estimate might be bogus because of the PAH emission at 8
% and 12 um
The observed dust temperature and stellar luminosity constrain the location of
the dust in the circumstellar environment. The distance of the dust from the
star at a given temperature also depends on the size of the dust grains and
the wavelength range of the light being absorbed. We can set limits on the
dust distance by first assuming the relation for large, blackbody grains,
\begin{equation}
\label{bbd}
R_{\rm BB}=2.32\times10^{-3}R_\star\left(\frac{T_\star}{T_{dust}}\right)^2,
\end{equation}
where $R_\star$ is the stellar radius in units of R$_\odot$ and $R_{\rm BB}$ is the dust
distance in au. The stellar radii for the stars are given in
Table~\ref{tab:class} and are taken from a second-order polynomial fit in $T$
to the values given in Appendix E of \citet{ostlie}. The distances calculated
with this model are given in Table~\ref{tab:ourdust} as the blackbody (BB)
distance. Grains that instead absorb light as blackbodies, but re-radiate as a
modified blackbody proportional to $\lambda^{-2}$, can be modeled using the
equation
\begin{equation}
\label{mbd}
R_{\rm MBB}=636^3\frac{\sqrt{L_\star}}{\lambda_0\,T_g^3},
\end{equation}
% possibly, either use p = q= 1.5 for the backman and paresce eqn, or just use
% the color temp eqn in gaustad and van buren - with backman and paresce, i'm
% guessing at the lambda_0, whereas gaustad and van buren's eqn uses grain
% size - backman and paresce's lambda_0 should be 2*pi*grain size, i think, so
% 10 um is probably too large
where $L_\star$ is in L$_\odot$, $R_{\rm BB}$ is in au, and
$\lambda_0$ is in \micron.  This equation is taken from Equation 2 of
\citet{back}, assuming $p=0$ and $q=2$. We have assumed $\lambda_0=1$
\micron. Note that this last parameter is poorly constrained and it
introduces significant uncertainty in $R_{\rm MBB}$. The distances
calculated with this model are listed in
Table~\ref{tab:ourdust}. \citet{smwy} note that in resolved debris
disks, the actual distances of the particles are at most 3 times the
distance predicted by the blackbody grain model, whereas in resolved
cirrus hot-spots, the dust distance is an order of magnitude -- or
more -- larger than that expected from the blackbody model, due to
stochastic heating of small grains \citep{neb}.

We also estimate the blow-out size for grains due to radiation
pressure in the absence of gas drag.  Grains with
$\beta=F_{rad}/F_{grav}>1$ will be blown out of the system by
radiation pressure. We use the equation for $\beta$ from
\citet{burns},
\begin{equation}
\beta=0.57Q_{pr}\frac{L/L_{\odot}}{M/M_{\odot}}
      \left(\frac{a}{\micron}\right)^{-1}
      \left(\frac{\delta}{\rm{g~cm}^{-3}}\right)^{-1},
\end{equation}
where $a$ is the size of the grain, $\delta$ is the density of the grain, and
$Q_{pr}$ is the average radiation pressure efficiency over the stellar spectrum
for a given grain size. Values for $L$ for the spectral types are given in
Table~\ref{tab:class} and were derived using
\begin{equation}
\frac{L}{L_\odot}=
                \left(\frac{R}{R_\odot}\right)^2\left(\frac{T}{T_\odot}\right)^4,
\end{equation}
where $T_\odot=5780$ K from Schmidt-Kaler \citep{allen}. Values for $M$ were
derived by fitting a second-order polynomial in $T$ in logarithmic space to
the values in Appendix E of \citet{ostlie} and are also given in
Table~\ref{tab:class}.
% VALUES FOR M ARE NOT CURRENTLY GIVEN IN TABLE 3 - JDS
We assume here $Q_{pr}=1$, and we use $\delta=3.3$ g cm$^{-3}$ from the
\citet{dustk} model. Setting $\beta=1$ gives the minimum grain size in a
circular orbit that can survive the radiation pressure of the host star.
Grains released from parent bodies in circular orbits, such as dust created in
a debris disk, however, can be ejected from the system for $\beta$ as low as
0.5.
% (table actually uses delta = 2 right now).

We then estimate the mass of dust needed to produce the observed
F$_{IR}$/F$_{Bol}$, based on the estimated blow-out size of the grains. This
mass (in units of g) is given by
% cite Jura et al, 1995?  morales et al cite Jura for this eqn - also, morales
% et al use 3 g/cm^3
\begin{equation}
M_{\rm bo}=\frac{16\pi}{3}\frac{F_{IR}}{F_{Bol}}\delta\, a_{min}\,R_{\rm BB}^{2},
\label{Mbo}
\end{equation}
where $a_{min}$ is the minimum grain size to
survive radiation pressure blow-out, in cm, and $R_{\rm BB}$ is the blackbody
distance in cm.

%-------------------------------------------------------------------------- 4.5
\subsection{Uncertainties}

We estimate the uncertainty on these parameters for the 8/24 pair of
excesses using the Monte Carlo method. For each of 1000 trials per
object, the color temperature is calculated, and that temperature is
used in the calculation of the dust mass, F$_{IR}$/F$_{Bol}$, and dust
distance. The uncertainties on the color temperature, dust mass,
F$_{IR}$/F$_{Bol}$, and dust distance are then set according to the
upper and lower values that bracket the central 67\% of the results
for each object. Uncertainties on the excess at 8 and 24~\micron\ are
determined via error propagation, assuming an uncertainty in slope $m$
of 0.05 for B stars and 0.01 for O stars. These ranges allow for
uncertainty in the spectral type. Because the O stars all have a
similar slope, the range is much smaller. We apply these uncertainties
to the excesses assuming a normal distribution about the value of the
excess before calculating the color temperature. The dust mass
calculation also includes an uncertainty in the distance to the stars,
which we model as a normal distribution centered at $61.1\pm6$ kpc,
assuming that the objects are most likely to be located near the
center of the SMC. The F$_{IR}$/F$_{Bol}$ calculation further requires
the uncertainty on the bolometric luminosity, assumed to be a uniform
distribution with a range set by the value for $\pm1$ spectral
sub-type. For the distance calculation, we assume uncertainties for
the stellar radius and stellar temperature uniformly distributed
between the values for $\pm1$ spectral
sub-type. Table~\ref{tab:ourdust} gives the dust parameters for those
objects in Table~\ref{tab:excess} that have a measurable excess at 8 $\mu$m,
and to which we give a secure main sequence spectral classification.

% i don't have uncertainties for MB distance, a_blowout, and blowout mass

%============================================================================ 5
\section{Discussion}
%-------------------------------------------------------------------------- 5.1
\subsection{Classical Be Stars}

Because the majority of our sample shows no optical emission lines, Be
star activity is likely not the main cause of the 24-\micron\ excess
in these O and B stars. It is possible, however, that a few of the
objects without emission lines could still be Be stars, as Be stars
are known to show variability in their Balmer emission. In a study of
45 Milky Way Be stars, \citet{mcswain} found 23 objects with variable
\ha\ emission, including six that changed from strong emission to
normal B-star absorption at \ha\ or vice versa over the course of just
two years. Of the original sample of 193 O and B stars with
24-\micron\ excess identified in the S$^3$MC survey, 12 were
previously catalogued as Be stars by \citet{meys}, and for those we
did not acquire new spectra. We identify here an additional 16 stars
with confirmed \ha\ emission and 7 stars with suspected weak \ha\
emission, for a total of 35 Be stars (36 if we include B\,097 in the
count). The remaining 88 stars in the spectroscopic sample do not
display evidence of OBe activity.
% CHECK the above statistics

To investigate the possibility of variability between the two states,
we look at the 4.5-\micron\ excesses, seen in Table~\ref{tab:excess},
because Be stars, both classical and Herbig Ae/Be, are expected to
show near-IR emission to some degree \citep[see,
e.g.,][]{beseds}. Indeed, we do see an excess ratio $>3\sigma$ above
the photospheric emission at 4.5 \micron\ for 11 of the 16 OBe stars
in our spectroscopic sample, as well as in 2 of the 7 ``OBe?'' stars.
This could indicate that the other 10 OBe and ``OBe?'' stars were in a
normal B-star phase during the IRAC observations, acquired in 2005
May. These excesses, however, are weak, at less than 4 times the
expected photospheric flux, suggesting that we may just be unable to
detect the excesses in our remaining spectroscopically confirmed OBe
stars because of the uncertainties ($\sim20-30\%$) in their
photospheric flux in our data. Adding further uncertainty is our use
of $J-$band data to scale the photosphere, because Be stars may
exhibit excesses even at wavelengths as short as 1.2~\micron. This can
be seen in the slopes of the dashed lines in the SEDs of B\,038,
B\,081, B\,085, B\,119, B\,140, and B\,190, for example, in
Figures~\ref{fig:be} and~\ref{fig:flat}. The dashed lines are fit to
$I-K$ and show shallower slopes than expected for the photospheric
emission.

We also find 12 stars (B\,023, B\,033, B\,047, B\,115, B\,128, B\,135,
B\,136, B\,142, B\,154, B\,156, B\,160, and B\,183) in
Table~\ref{tab:excess} that do not show optical emission lines (see
Table~\ref{tab:spec}) but do have an excess ratio at 4.5~\micron\
significant at the $3\sigma$ level. The 4.5-\micron\ excesses are also
weak, with only B\,136 showing an excess/photospheric ratio greater
than 1. This excess could indicate that these stars were in a Be state
at the time of the IRAC photometry but settled into a normal B-star
state by the time the spectra were acquired. If we assume that all of
the 4.5-\micron\ excesses are due to Be-star behavior, the fraction of
transient Be-stars (counting as transient those OBe and ``OBe?'' 
optically classified stars without 4.5~\micron\ excess, plus these 12
stars with 4.5~\micron\ excess but no hint of H$\alpha$ emission in
their spectra) is 22/47, or about 47\% of the total Be-star population
in the sample. This is on par with the fraction seen in the study of
Galactic Be-stars by \citet{mcswain}. Therefore it seems unlikely that the
remainder of our stars that show no Be activity in the photometry and
spectra are primarily dormant Be-stars.

%-------------------------------------------------------------------------- 5.2
\subsection{Herbig Ae/Be Stars}

Because we do not know the ages of the Be stars in our sample, it is
difficult to determine to which of the Be-star categories they
belong. None of the Be stars we identified spectroscopically have the
large excesses at $\lambda\gtrsim 1$ \micron\ that are typical of the
hot dust in Herbig Ae/Be objects. \citet{hill}, however, identify in
the Galaxy a set of objects with near-IR profiles that resemble
classical Be stars, but are apparently young stars, located near
reflection nebulosity. This location is more typical of Herbig Ae/Be
stars. \citet{hill2} suggest that these objects, which they label
Group~III, are actually Herbig Ae/Be stars that have an optically thin
circumstellar disk, rather than the optically thick disk more typical
of the Herbig Ae/Be class.

The weak 4.5-\micron\ excesses of the Be stars detected in our spectroscopic
study resemble the near-IR profiles of classical Be stars as well as these
Group~III objects. The stars in our sample are located within regions of
recent star formation, suggesting an age of $\lesssim 10$~Myr. Unfortunately,
10~Myr is the approximate age at which classical Be stars begin to appear in
clusters \citep{fab}, and so we cannot be certain whether these objects are
young stars or older, established main-sequence stars without more accurate
age determinations. The bulk of the dusty B-star sample, with neither optical
emission lines nor a near-IR excess, are unlikely to be Herbig Ae/Be stars.

%-------------------------------------------------------------------------- 5.3
\subsection{Cirrus Hot-Spots Comparison}

\citet{back} note that to be certain that debris disks are the source of the
excess infrared emission, spatially resolved observations of the disk
are needed, because hot spots in the ISM can mimic the color
temperatures and fractional luminosities of debris disks. The
morphology of the dust is the best clue as to whether the heated dust
is physically associated with the star in a disk, or if it is simply a
concentration of ISM material close to the star. Unfortunately, at the
distance of the SMC, debris disks cannot be resolved with any
available facilities, and ISM hot spots are below the resolution of
the {\it Spitzer} data, especially with a 6\arcsec\ resolution at
24~\micron.

Since we lack resolved images of the dust emission, we instead compare the
dust properties of the SMC sample to the Galactic cirrus hot-spot sample of
\citet{gaust}. A broader comparison using {\em WISE} data is carried out 
in Paper II. The \citet{gaust} sample was selected by looking for {\it
IRAS} excesses at 60~\micron\ with color temperatures above the
$\sim25$~K temperature of the general Galactic cirrus and peaking at
the location of a star, with $F_{IR}/F_{Bol}<0.1$ to exclude \hii\
regions, and spatially extended emission that is more extended in
longer wavelengths, indicating that the temperature declines with
distance from the star. We note that in cases where the reported {\em
IRAS} 25-\micron\ flux is negative, the color temperatures cannot be
calculated.

Whether this cirrus hot-spot sample in fact consists solely of cirrus
hot-spots is, however, uncertain. \citet{gaust} found that in some
cases, the hot spots have 12-\micron\ emission that is brighter than
the 25-\micron\ emission, indicating a hotter dust component than the
expected $\sim70$~K temperature of illuminated cirrus. Indeed, the
color temperatures we find for the hot-spot sample from the 12/25
flux-ratios are all above 100~K.  Furthermore, \citet{neb} used
resolved optical images of nebulosity around three B stars that appear
in both the \citet{gaust} and \citet{back} samples to model the IR
emission from the objects. The authors concluded that, in all three
cases, the ISM model cannot explain the 12- and 25-\micron\ emission,
requiring a source of warmer dust. The coronograph used to block the
starlight for their optical images unfortunately obscures the central
100--1000~au around the star, where a circumstellar disk would be
located. The warmer dust could therefore be provided by a population
of either blackbody grains in a close-in circumstellar disk, or very
small grains at larger distances that are heated stochastically to
higher temperatures than equilibrium. In cases where the stellar
velocity is large compared to the local ISM (such as runaway stars), a
bow-shock could further heat the dust in addition to stellar
irradiation.  Distinguishing between these scenarios would require
imaging at higher resolution, which we discuss further in Paper
II. The 12-\micron\ IRAS band also contains a PAH emission feature,
offering another possible explanation for the excess flux at that
wavelength.  Illustrating the difficulty of distinguishing disks
from hot-spots using SED information alone, even for nearby objects,
\citet{neb} note that 31 of the 34 B-type stars in
the \citet{back} table of debris disk candidates or Vega-like stars,
appear in the
\citet{gaust} cirrus hot-spot list.

The full hot-spot sample contains stars of spectral types O6 through
B9 of all luminosity classes, but we exclude from the comparison any
objects marked in the \citet{gaust} table as emission line stars or
stars of luminosity class higher than V, to match the characteristics
of the SMC dusty OB stars. We further limit the hot-spot sample to
those objects that have excesses, if scaled to the SMC distance of
61.1 kpc, that we could detect at the 5-$\sigma$ level or better at
24~\micron\ in the S$^3$MC survey (215~$\mu$Jy;
\citealt{bolatto}). This sub-sample contains 18 stars. We then
calculated the dust parameters for this sub-set of the sample, using
the 12/25-\micron\ combination in place of 8/24-\micron, and these
values are listed in Table~\ref{tab:gaudust}. Stars flagged in
Table~\ref{tab:gaudust} as ``known reflection nebula'' are noted in
\citet{gaust} to have extended emission that is brighter in the blue than in
the red on the Palomar Sky Survey or ESO Sky Survey plates.
% I have no O7 in gaustad, so removed it from the paramhists plots....

\begin{figure}
\epsscale{1.0}
\plotone{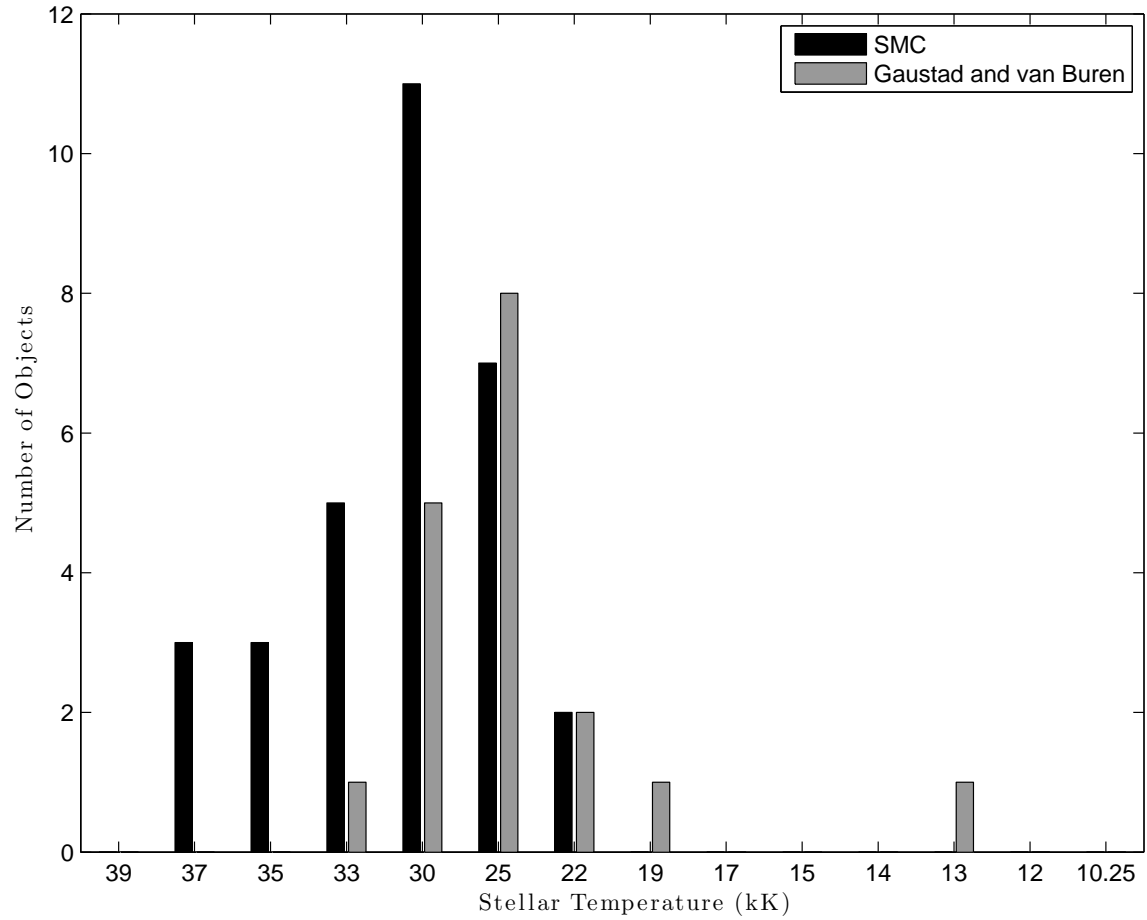}
\caption{Total number of objects in each stellar temperature bin for which the
8/24 or 12/25 dust temperature has been calculated in Tables~\ref{tab:ourdust}
and \ref{tab:gaudust}. The SMC sample is shown in black, while the
\citet{gaust} sample of Galactic cirrus hot-spots is shown in gray. There are
31 SMC stars and 18 hot spots.}
\label{fig:maxhist}
\end{figure}

The comparison SMC sample contains a total of 31 stars (c.f.,
Table~\ref{tab:ourdust}), and it is constituted by normal main
sequence stars (no OBe, ``OBe?,'' or Em classification) that have a
measurable 8~\micron\ emission excess, in order to determine the
temperature of the dust.  While the distributions in both samples over
spectral types peak at a similar type -- B0 in the SMC vs.\ B1 in the
hot-spot sub-sample -- the SMC sample extends more towards earlier
types while the hot-spot sub-sample extends towards later types. This
is reflected in the distributions over stellar effective temperature,
shown in Figure~\ref{fig:maxhist}.

\begin{figure}
\epsscale{1.0}
\plotone{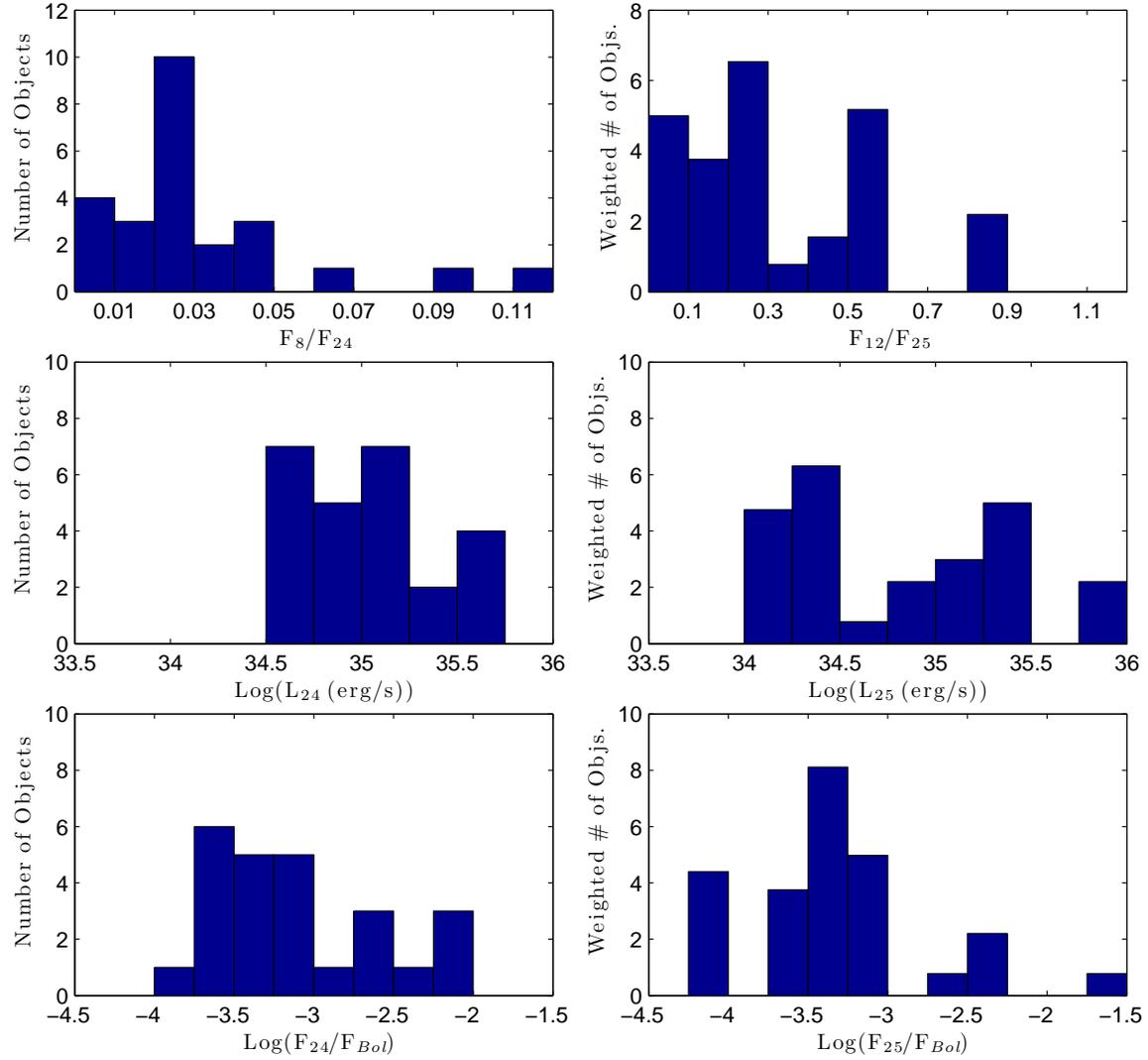}
\caption{Histograms of the observable parameters displayed in
Figure~\ref{fig:vstemp}. On the left are the histograms for the SMC objects,
using only those with spectral type O9--B2. On the right are the histograms
for the \citet{gaust} sample, limited to only those objects that would have a
24~\micron\ flux greater than $\sim215$~$\mu$Jy at the distance of the SMC.
This cutoff is chosen since it is the 5-$\sigma$ flux limit of the S$^3$MC
survey at 24~\micron.}
\label{fig:maxobs}
\end{figure}

\begin{figure}
\epsscale{1.0}
\plotone{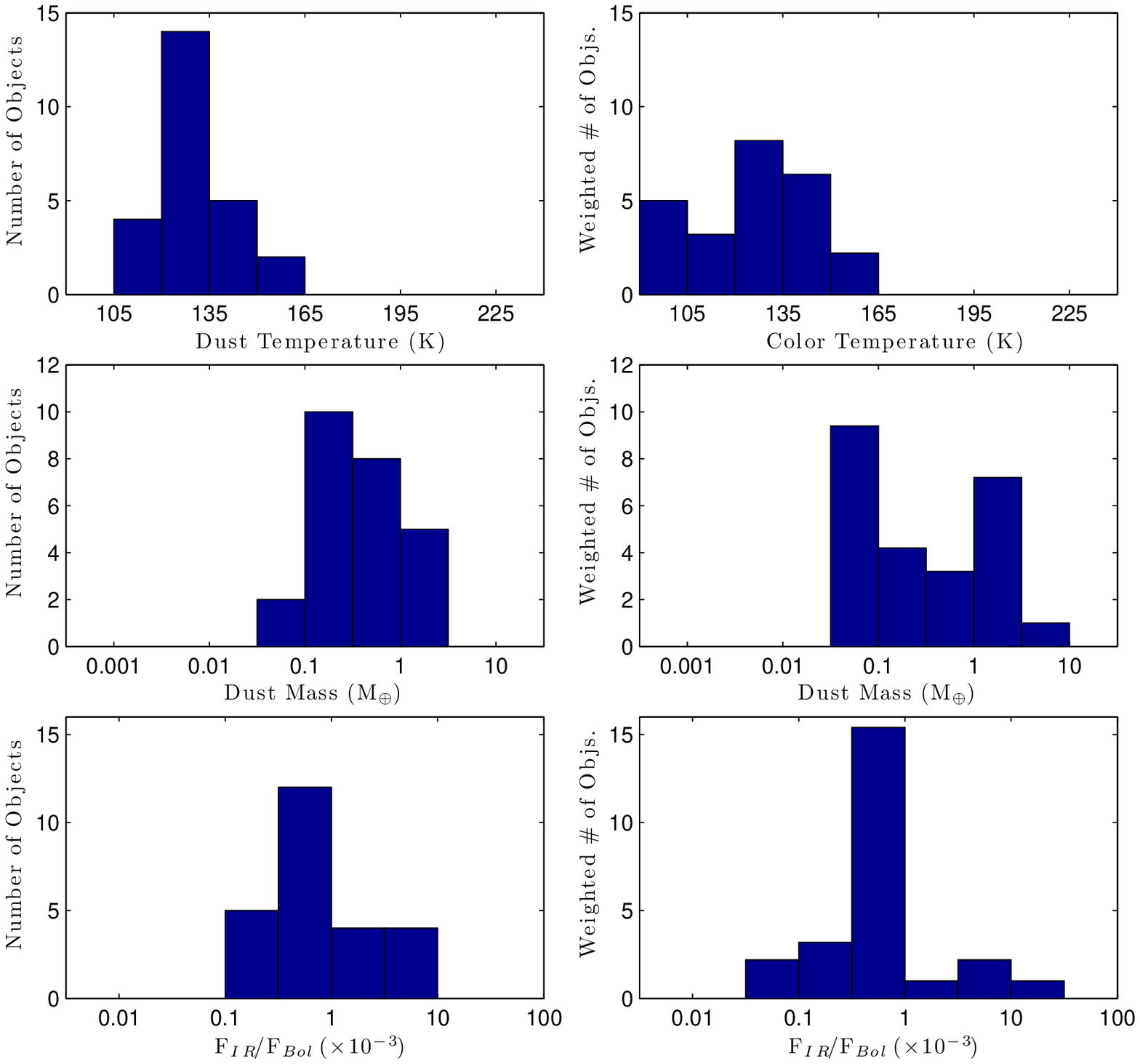}
\caption{Histograms of the physical parameters in Figure~\ref{fig:vstemp},
derived from the observable parameters in Figure~\ref{fig:maxobs}. On the left
are the histograms for the SMC objects, using only those with spectral type
O9--B2. On the right are the histograms for the \citet{gaust} sample, limited
to only those objects that would have a 24~\micron\ flux greater than
$\sim215$~$\mu$Jy at the distance of the SMC. This cutoff is chosen since it
is the 5-$\sigma$ flux limit of the S$^3$MC survey at 24~\micron.}
\label{fig:maxdust}
\end{figure}

\begin{figure}
\epsscale{1.0}
\plotone{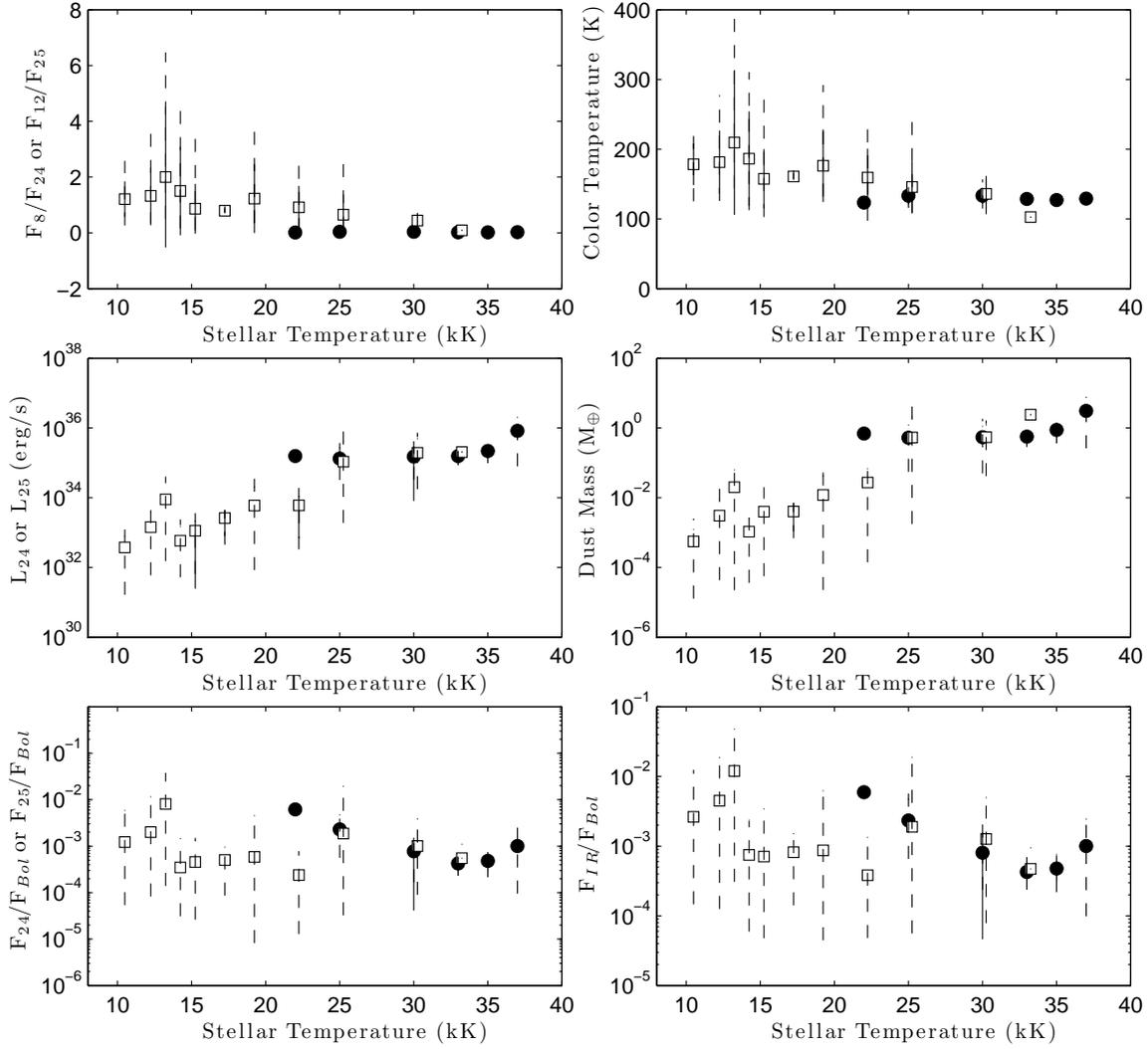}
\caption{Average properties for the SMC dusty OB stars (circles) and the
\citet{gaust} cirrus hot spots (squares) as a function of stellar temperature.
On the left are the observable quantities from which the physical parameters
on the right are derived. The solid bars on each point represent the standard
deviation, and the dashed lines connect the maximum and minimum value for each
stellar temperature. The \citet{gaust} points are offset in stellar
temperature by +250~K for clarity. The spectral types O7--B2 correspond to
stellar temperatures of 37,000 to 22,000~K. Points without bars only have one
or two objects included in the average, except in the case of the
F$_{IR}$/F$_{Bol}$ panel. The standard deviation for this panel has a negative
lower bound for all of the hot spots and for O7 (37,000~K), B0 (30,000~K), and
B1 (25,000~K) in the SMC sample.}
\label{fig:vstemp}
\end{figure}

In Figure~\ref{fig:maxobs} we show histograms of the observable
parameters, F$_8$/F$_{24}$ or F$_{12}$/F$_{25}$, L$_{24}$ or L$_{25}$,
and $F_{24}$/F$_{Bol}$ or $F_{25}$/F$_{Bol}$. The IR luminosities and
fractional IR luminosities are similar; the difference in the IR ratio
is largely driven by the use of the 8- vs.\ 12-\micron\ bands. In
Figure~\ref{fig:maxdust} we show histograms of the color temperature,
mass, and fractional luminosity, where the samples are restricted to
spectral types O9--B2, since both samples have at least one star of
each of these spectral types. The histograms for the hot spots in
Figures~\ref{fig:maxobs} and \ref{fig:maxdust} are created by
weighting the number of objects in that sample so that the total
number of hot spots in a spectral type bin equals the total number of
objects in the bin for the SMC sample. In this manner the weighting
controls for the different spectral makeup of the
two samples. Figure~\ref{fig:vstemp} shows the averages of each of the
observable parameters and associated dust parameters against stellar
temperature, without the restrictions on visibility at the SMC
distance that are used in the histograms.

The distributions over dust temperature are very similar for both
samples. they both have similar mean temperatures in the 125--130~K
range, although the hot-spot sample also displays a broader range in
dust temperatures, 97--226~K compared to 116--164~K for the SMC
sample. The hottest color temperatures in the hot-spot sample in
Figure~\ref{fig:maxobs} result from the coolest and most predominant
spectral type, B2. The trend toward higher color temperatures in the
hot-spot sample is possibly due to PAH contamination. The 12-\micron\
PAH emission feature remains relatively unchanged over a broad range
of radiation field intensities, but the thermal emission at
25~\micron\ changes dramatically (e.g., Figure 13 in
\citealt{pahs}). Thus PAHs around cooler stars would create higher
color temperatures than around hotter stars; F$_{12}$/F$_{25}$ would
be artificially high as a result of similar amounts of PAH emission at
12~\micron\ but reduced thermal emission at 25~\micron.  This is
likely a lesser effect in the SMC, where PAH emission is weak
\citep[e.g.][]{smcpah}.

% Say something about the F24 vs F25 histograms and averages, since mass
% assumes color temp is real temp.
The distributions over dust mass are also similar for both samples, though
again the distribution appears broader (both towards higher and lower dust
masses) in the hot-spot sample. The SMC sample may be devoid of examples of
lower dust mass as these would not have been detected. The average mass at B0
and B1, where there is the greatest number of objects in both samples, is
similar for each sample (Fig.~\ref{fig:vstemp}), but the average dust mass in
the SMC sample is higher at B2, and lower for O7--O9, compared to the cirrus
hot-spots. We see a trend in the hot-spots of lower dust mass for cooler
stars, which is reasonable given that a hotter, more massive star could
illuminate larger regions of the ISM. We do not see this dependence on the
stellar temperature for the SMC objects, although the smaller temperature range
of the SMC sample would make such a trend perhaps difficult to detect.

% The caveat here is that Fir/Fbol is derived using color temp as a real dust
% temp, so compare also the F24/Fbol and F25/Fbol plots.
At first glance, the distributions over $F_{IR}/F_{Bol}$ do also
appear similar between the two samples. On closer inspection, the one
for the hot-spots is clearly broader. The average
$F_{IR}/F_{Bol}=4\times10^{-4}$ for the hot-spot sample is smaller
than that of the SMC, which is $F_{IR}/F_{Bol}=1.5\times10^{-3}$, 
although this is somewhat misleading as the averages are skewed by the
extremes in the sample.
% misleading as slightly different value leads to similar numbers:
% Only 4 of the 39 hot spots have $F_{IR}/F_{Bol}>5\times10^{-4}$, while 21 of
% the 31 SMC objects do.
The derived $F_{IR}/F_{Bol}$ for both samples do fall within the typical
$10^{-4}$ to $10^{-2}$ range of debris disks \citep{chen06}.

While we noted minor differences in the distributions over dust temperature,
dust mass and $F_{IR}/F_{Bol}$ between the SMC and hot-spots samples, these can
be explained by a combination of small-number statistics, selection bias, and
the opposite sense of skewedness of the distributions over spectral type.
Indeed, Figure~\ref{fig:vstemp} shows that the trends -- or absence thereof --
and mean values are consistent between both samples if the breakdown over
spectral types is taken into account.

Interestingly, despite the low dust-to-gas ratio in the SMC
\citep[about $1/7-1/10$ that of the Milky Way;][]{dusttogas}, the SMC
objects show dust masses similar to the Galactic hot-spots if common
selection thresholds are applied. This suggests that, if they are
hot-spots, the mass of gas being illuminated is $7-10$ times larger in
the SMC than in the Milky Way. Assuming similar gas densities, this
suggests that the typical sizes of hot-spots would be $\sim2$ larger in
the SMC than in the Milky Way.
%It does beg the question, then, as to why the SMC objects should show such
%large masses, similar to the Galactic hot-spots if the common selection
%thresholds are applied, as the SMC has a dust-to-gas ratio of about $1/7-1/10$ that
%of the Milky Way \citep{dusttogas}. This suggests that the SMC objects are not
%typical ISM hot-spots, as it would require a much larger mass of illuminated
%interstellar gas in the SMC to produce such large amounts of emitting
%dust. Perhaps those extreme objects are simply not included in the
%\citet{gaust} sample -- after all the {\it Spitzer} surveys of the SMC sampled
%an entire galaxy whereas the \citet{gaust} survey only sampled a small portion
%of the Milky Way disc (with all the usual complications of confused
%sightlines). The larger beamsize of the SMC survey may also have included more
%emission than a beamsize of the same physical size as used in the
%\citet{gaust} survey would have done. 

%-------------------------------------------------------------------------- 5.4
\subsection{Bow-shocks and Runaway Stars}

Larger dust masses, warmer dust and thus brighter emission may arise
from bow-shocks ahead of runaway stars, or stars being over-run by
expanding interstellar bubbles. Stars with stronger stellar winds --
i.e., of earlier type and/or at later evolutionary stages -- and/or
larger relative velocity with respect to the local ISM sweep up more
mass; larger relative velocities also result in stronger shocks and
hence greater heating. \citet{gvaramadze} present a sample of 12
candidate runaway stars with bowshocks in the SMC. Curiously, there is
no overlap whatsoever between their sample and ours.

\begin{figure}
\epsscale{1.0}
\plotone{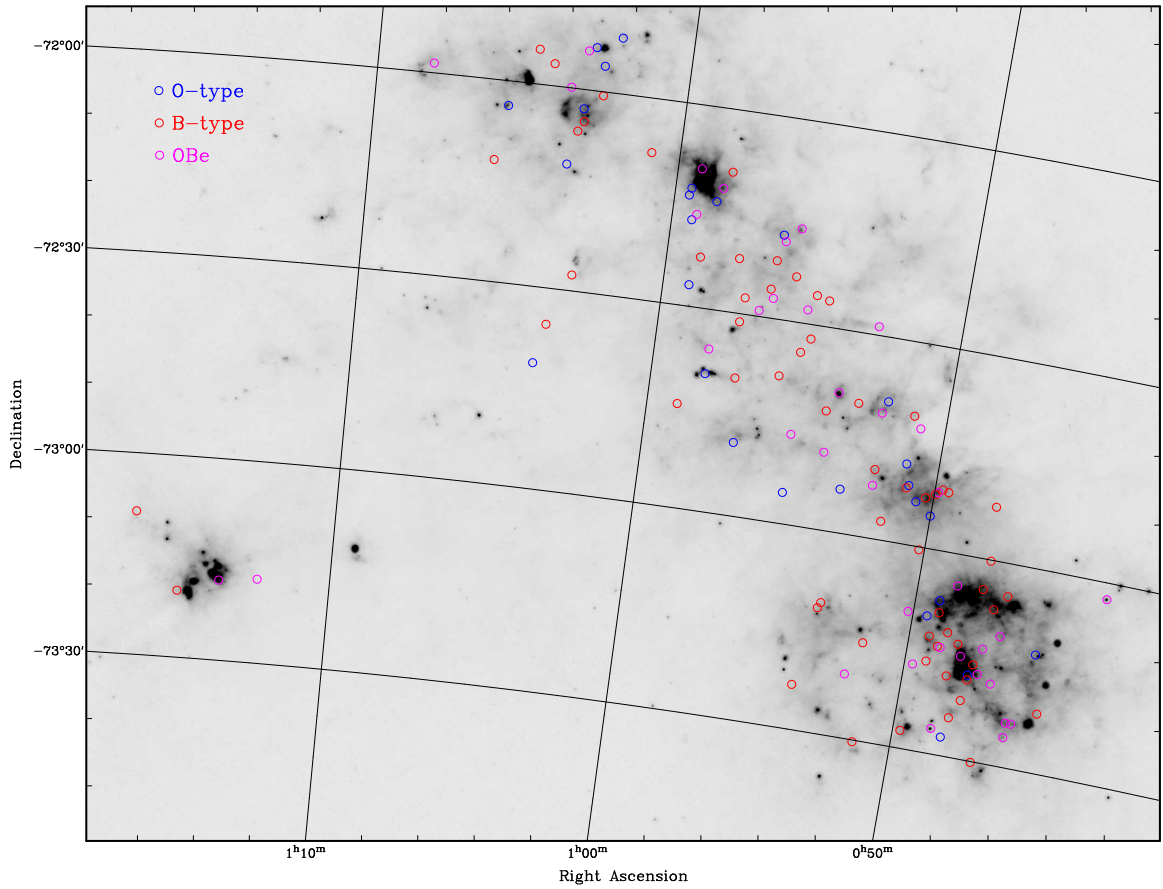}
\caption{Map of the 70-\micron\ emission (SAGE-SMC, \citealt{gordon11}) with
overlain the dusty O-type stars (blue), B-type stars (red) and other, mainly
OB-type emission-line stars (magenta). The latter tend towards the peak of
70-\micron\ emission but the B-type and even O-type stars often stray away
from prominent star-forming regions suggesting some may be runaways (not
necessarily with a bow-shock).}
\label{fig:map}
\end{figure}

The distribution of our sample on the sky (Figure~\ref{fig:map}) seems to
indicate that at least a portion of them are found outside of sites of
vigorous star formation. While the late-B stars may be old enough to have
migrated from their birth-sites or for their natal clouds to have dispersed,
the fact that some of the O-type stars are also found outside star-forming
regions suggests they may be runaways. One of our ``normal'' OB stars, B\,064
is in fact the primary in an X-ray binary\footnote{The other primary in an
X-ray binary in our sample, B\,085, is a Be star and therefore excluded here
from further analysis}, and it is likely that the binary system will have
received a high peculiar velocity when the neutron star's progenitor star
exploded as a core-collapse supernova -- see \citet{kaper} for the
proto-typical example of a bow-shock accompanied runaway X-ray binary,
Vela\,X-1.

We also note that the structure of the ISM in the SMC differs in important
ways from that in the Milky Way disk; the SMC is dominated by expanding
bubbles \citep{lister} while gas in the Milky Way disc is more strongly
entrained and collected by the recurrence of a spiral density wave. This could
affect the ubiquity and properties of bow-shocks; for instance stars in the
SMC might be over-run by expanding shells \citep[see][for such
mechanisms operating within the Large Magellanic Cloud]{dawson}, resulting in large
relative velocities between stars and ISM without the requirement for a
``kick'' velocity of the star.

While proper motions relative to the systemic motion of the SMC are
too small to be measured (1 mas yr$^{-1}$ $\approx$ 300 km s$^{-1}$),
radial velocities can be determined with much better accuracy. Our own
spectroscopy, however, has neither the spectral resolving power nor
velocity calibration accuracy required for such measurements. In
Table~\ref{tab:spec} we list the radial velocities measured for stars
in our sample by \citet{evans06}, \citet{martayan07} and
\citet{evans08}. We also determined the velocity of the peak in the \hi\
emission around the location of the star in question, from inspection of the
datacube produced by \citet{stanimirovic99}. This differs from the
intensity-weighted velocity map produced by \citet{stanimirovic04}, because we
are interested in the most likely velocity difference between a star and the
densest parts of the ISM. Often, more than one strong peak is seen in the \hi\
data; in that case we list both. Of the 21 stars with
radial velocity measurements, most stellar kinematics coincide with that of a
strong \hi\ peak. B\,116, B\,124 (and perhaps B\,119) fall in between \hi\
peaks, suggesting they might reside in the middle of a shell expanding at
$\sim18$--20 km s$^{-1}$. But two O7 stars, B\,142 and especially B\,145, have
large velocity differences of $\sim35$ and 75 km s$^{-1}$ with respect to the
nearest (in velocity) \hi\ peak. No uncertainties were quoted for the stellar
radial velocities for these stars, but measurements for other stars from the
same work (with the caveat that those were all later spectral type) often
agreed with \hi\ kinematics, which suggests typical errors not much larger
than $\sim10$ km s$^{-1}$. We thus suggest that these two O7 stars may be
runaways -- they are located just outside of the brightest \hii\ region in the
SMC, LHA\,115-N\,66 containing the massive O-star cluster NGC\,346, lending
further support to their large space velocities.

%-------------------------------------------------------------------------- 5.5
\subsection{Disks}
% Jacco raised the question of whether debris disks have emission lines as
% well - the Uzpen et al paper does indeed have debris disks that also show
% halpha emission - if they show balmer emission, no Ca II emission, and > 3
% sigma excess at 24 micron, once the free-free component has been removed,
% then they are TRANSITION disks - problem is that the Ca II lines are
% indistinguishable from the Paschen lines at our resolution - if there IS Ca
% II emission, then it's a HAe/Be star (see Figure 18 in their paper)
%
% Stark's paper on 51 Oph is also a debris disk (Beta Pic analog) which shows
% Halpha emission

Statistics for circumstellar disk detections around stars with
spectral types as early as the SMC sample are difficult to come
by. Few surveys exist of stars younger than $10$ Myr, which
encompasses most, if not all, of the main-sequence lifetimes of these
late-O and early-B stars, and those surveys that do sample the
appropriate age range still contain very few stars of these types
\citep[see][and references therein]{carp}.

Our sample of 87 objects with normal O- or B-star main sequence
spectra in the SMC shows excesses at 24~\micron\ that are larger than
those of typical debris disks, in studies of somewhat less massive
(late-B and A-type) stars in the Milky Way. \citet{carp} set a limit,
based on the properties of their sample of B7--A9 stars, for the ratio
of observed flux to expected photospheric flux at 24~\micron\ of
$\lesssim 100$ for debris disks (but only a ratio of $\lesssim 5$ for
solar-type stars). Any object with a larger ratio is considered to be
a primordial disk. \citet{hern6} define objects with $K-[24] \gtrsim
5$ mag as Herbig Ae/Be stars, while objects below this limit are
debris disks. \citet{hern6} also define any object with $2.3
\lesssim K-[24] \lesssim 5$ mag as a massive debris disk. The SMC objects have
observed flux to expected photospheric ratios $\gtrsim 70$, with only 6
objects being less than 100 and a further 7 objects being within 1~$\sigma$ of
100. The SMC objects also have $K-[24] \gtrsim 3.4$ mag, with only 11 objects
below the limit of 5. Of those 11, we identify 7 as OBe stars.

The studies of \citet{hern6} and \citet{carp} only consider
circumstellar matter around intermediate and low mass stars, so it is
possible that these dividing limits around the more massive stars in
our SMC sample should be higher. \citet{carp}, however, specifically
excluded earlier type stars from consideration because, aside from
being few in number, they may show a severely reduced dust luminosity
due to the loss of small grains to radiation pressure. We find blowout
grain sizes for the SMC objects on the order of 1 mm, in the absence
of gas drag (see Table~\ref{tab:ourdust}). The smallest dust grains
are the most efficient (per unit mass) at capturing and reprocessing
UV and optical radiation into infrared. This suggest that, if blowout
is significant, $F_{IR}/F_{Bol}$ started much larger than we currently
measure in our sample. 
%Also, we note the high frequency of
%binaries among our SMC sample, and binarity may well lead to dynamical
%disruption of circumstellar disks.
% need to discuss the results of the mass derived from the blowout size - that
% mass is the mass of large grains needed to explain the Fir/Fbol. The masses
% needed to explain the Fir/Fbol are quite a bit larger than the masses from
% the dust model... but the dust model implies grains smaller than 1mm?? If
% there's gas in the system, that could trap the smaller grains. Could be
% cooler gas a little further out, so we don't see the Halpha emission, but do
% see emission from small grains? Morales et al do this calculation (but using
% a narrower wavelength range for Fir) and get about 10^-3 LUNAR masses for
% their objects, which are A/lateB stars

\citet{hern6} suggest two scenarios to explain massive circumstellar 
dust disks. They could be the result of collisions between two large
planetesimals, greater than 1000~km, or they could be explained as a
transition phase from Herbig Ae/Be star to debris disks, in which the
disk still contains a significant fraction of primordial dust. The
small number of these objects detected in the SMC sample suggests that
they could be in a short-lived phase, and perhaps constitute
transition disks remnants from the stellar accretion. In the original
S$^3$MC sample, the 24-\micron\ excess objects were 5\% of the total
number of objects that met the requirements for normal late-O and
early-B stars in the specific magnitude and color cuts
\citep{bolatto}. Counting only the stars that have normal O- and B-star spectra,
which is 87 out of the 193 objects in the original list, and also assuming
that all of these objects are debris disks, rather than ISM hot spots or
quiescent Be stars, gives a detection fraction of only 2.2\%. Including the 45
objects for which we have no spectra to judge their emission-line star status
gives a maximum of 3.4\%. If the possible transition object that \citet{hern6}
detect is truly in this transition phase, that would give a detection fraction
of 3\% for transition objects in the 5~Myr cluster in which it lies. This
detection fraction is consistent with that seen in the SMC sample.

Finally, it is worth noting that comparison with literature
classifications shows that a significant fraction of our SMC sample
consists of binaries. The effect of binary systems on the formation of
circumstellar disks is unclear. One might expect the disks to be dynamically
disrupted, although circumbinary planets have been recently discovered by 
{\em Kepler} \citep{doyle,welsh,orosz1,orosz2,schwamb}.

%============================================================================ 6
\section{Summary and Conclusions}
% has not been edited to consider PAHs or the observable parameter figures

We obtained long-slit spectra of 125 of the 193 dusty OB stars in the
S$^3$MC survey that show large excess emission at 24~\micron. These
new observations cover a significant fraction of the sample, make
possible accurate sky removal in the complex regions where these stars
live, and allow us to draw robust conclusions on the statistics of
different contributors to the sample.

We use these spectra to classify the stars and look for the signatures
of the OBe activity.  We find 87 objects that lack emission lines and
appear to be normal, main sequence late-O/early-B main-sequence
stars. For these stars, we use our spectral classification to estimate
their luminosity, and calculate the dust temperature, dust mass,
F$_{IR}$/F$_{Bol}$, and equilibrium dust distance from the host
star. We discuss the spectral and photometric properties of our sample
in relation to several possible scenarios: classical Be stars, massive
analogs of Herbig Ae/Be stars, hot-spots in the interstellar medium
(possibly with contributions of bow-shocks), and disks. The results
of this study are:

\begin{itemize}
\item[$\bullet$]{We identify 17 stars in the dusty O and B stars sample as 
OBe stars because of clear \ha\ emission in their spectra (this count
includes B\,094, the object with forbidden nitrogen emission). A
further 7 stars exhibit weaker than expected \ha\ absorption and are
therefore classified as possible OBe stars -- one of these is an
O6\,Iaf star showing \ion{He}{2} 4686 \AA\ emission. We also identify
12 stars that show no \ha\ emission in their spectra but do show
4.5~\micron\ excess in the photometry that could be indicative of
transient Be-star activity. If these 12 stars are dormant OBe stars,
along with the 11 spectroscopically identified OBe stars that show no
excess at 4.5~\micron, the transient total OBe star fraction would be
$\sim47$\%. This fraction is consistent with the study of transient Be
star activity by \citet{mcswain}. Thus it seems unlikely that the
remaining 75 stars, with neither 4.5~\micron\ excesses nor \ha\
emission, are OBe stars that happened to be in normal OB star states
during both the photometric and spectroscopic observations.}
% I had cut this: If the O and B stars in the SMC with 24~\micron\ excess are
% Be stars with sometimes absent emission, it would require that an
% unrealistically large fraction (75 to 86 \%) of the objects were in a normal
% B star phase during the acquisition of the optical spectra.
\item[$\bullet$]{
%The weak excess in the near-IR of our sample of emission-line
%stars resembles the free--free emission seen in classical Be
%stars. 
Our dusty OB stars are generally located near regions of recent star
formation, similar to Herbig Ae/Be type stars, and in that respect
resemble the Group III Herbig Ae/Be objects in \citet{hill}. Our
objects have ages $\lesssim 10$~Myr. Note that classical Be-star activity
begins to appear in clusters around $10$~Myr of age, once again suggesting that
the majority of them cannot be explained by excretion disks with a dusty component.}
\item[$\bullet$]{The SMC sample exhibits broadly similar characteristics when compared to the cirrus hot-spot sample of \citet{gaust}.
The hot-spot sample shows
trends for decreasing dust temperature and increasing dust mass for hotter
stars, which the SMC sample does not. However, these differences likely arise
from the different distributions over spectral type, and perhaps stronger contamination of the broadband colors by PAHs in the Milky Way. 
%The large(r) dust masses
%for the SMC objects are surprising given the lower dust-to-gas ratio at the
%low metallicity of the SMC. 
We show evidence that at least some of the stars
in the SMC sample are runaways, including one X-ray binary; heating in a
bow-shock could add to the radiative heating of the dust.}
% I had cut this out: However, this could be an artifact of the small number
% of objects in the stellar temperature bins for the B2 (22,000 K) spectral
% type in the SMC sample and the O7-9 (37,000 K to 33,000 K) spectral types in
% the hot spot sample. The average values for the two samples do overlap
% within a standard deviation for B0 (30,000 K) and B1 (25,000 K), which are
% the spectral types that have the best numbers of objects in both samples.
\item[$\bullet$]{The SMC sample of objects show higher 24~\micron\ excesses than
typical debris disks around intermediate mass stars in the Milky
Way. There are not enough debris disks known around comparably massive
stars, however, to indicate what is typical for an early-B or late-O
debris disk. Alternatively, if the stars are young, we may be
observing the remnants of the accretion process --- analogs to transition
disks around massive stars. The detection fraction of these objects
when compared to stars with similar optical colors and luminosities 
is 2.2\%--3.5\%, which is consistent with the expected
short lifetime of a transition phase. }
%On the other hand, the small
%detection fraction could just be an indication that we are seeing only
%the brightest, and thus most massive, disks in a much larger
%population.}
% I had cut this: The spectra and analysis presented here are but a step in
% uncovering the origin of the excess in the dusty O and B stars in the SMC.
% The majority of the objects do not appear to be emission line stars.
% Distinguishing between local heating of the ISM and heating of a
% circumstellar debris disk left over from planet formation is more difficult,
% as the dust parameters for the SMC sample are consistent both with Galactic
% debris disks as well as Galactic cirrus hotspots.
\end{itemize}
\vspace{0pt}
We analyze and discuss additional observations of the SMC dusty OB
sample, as well as further comparison with a sample of massive stars
in the Galaxy compiled using {\em WISE} photometry, in Paper II.

\acknowledgements{Based on observations obtained at the European Southern
Observatory, programme 079.C-0485 (PI J.Th.van Loon). A. D. B. wishes
to acknowledge partial support from a CAREER grant NSF- AST0955836,
JPL-1433884, and from a Research Corporation for Science Advancement
Cottrell Scholar award. This research has made use of the SIMBAD
database, operated at CDS, Strasbourg, France. A. D. B. thanks
Paul Kalas and Eugene Chiang for discussions that spurred the initial
stages of this research.}

%===============================================================================

\clearpage

%====================================================================== Figures

% FIGURE 1

% FIGURE 2

\clearpage

% FIGURE 3

\clearpage

% FIGURE 4

\clearpage

% FIGURE 5

\clearpage

% FIGURE 6

\clearpage

% FIGURE 7

\clearpage

% FIGURE 8

\clearpage

% FIGURE 9

\clearpage

% FIGURE 10

% FIGURE 11

\clearpage

% FIGURE 12

% FIGURE 11

\clearpage

% FIGURE 14

\clearpage

% FIGURE 15

\clearpage

% FIGURE 16

\clearpage

% FIGURE 17

\clearpage

% FIGURE 18

\clearpage

%======================================================================= Tables
% NOTE!!!!  ellipsis is \ldots (that's L)
%
% NOTE: need to separate B0 from B0.5 (have fixed 8-$>$8.5) - had rounded all
% spec types before to O8 and B0, because I had trouble reading data into IDL
% only 1000 K diff for O8 to O8.5, but is 3000 K for B0 to B0.5

%%%%% TABLE 1 - phottab.txt
\LongTables
\begin{deluxetable}{lccccccccc}
\tabletypesize{\tiny}
\tablewidth{0pt}
\tablecolumns{10}
\tablecaption{SMC Dusty O and B Stars: Optical and 2MASS
Data\label{tab:phottab}}
\tablehead{
\colhead{ID No.} &
\colhead{Name} &
\colhead{RA (J2000)} &
\colhead{Dec (J2000)} &
\colhead{B\tablenotemark{a}} &
\colhead{V\tablenotemark{a}} &
\colhead{I\tablenotemark{a}} &
\colhead{J} &
\colhead{H} &
\colhead{K} \\
&
&
\colhead{(deg)} &
\colhead{(deg)} &
\colhead{($\mu$Jy)} &
\colhead{($\mu$Jy)} &
\colhead{($\mu$Jy)} &
\colhead{($\mu$Jy)} &
\colhead{($\mu$Jy)} &
\colhead{($\mu$Jy)} \\
}
\startdata
% 27-Oct-2010 15:41:02
B001 & J004336.56$-$730226.80 & 10.90234 & $-$73.04078 &  2078.39 & 1585.23 &  918.67 & 1088.65\tablenotemark{a} &  564.29\tablenotemark{a} &  827.17\tablenotemark{a} \\
B003 & J004511.14$-$732111.95 & 11.29644 & $-$73.35332 &  1909.53 & 1424.6  &  780.47 &  387.0                   &  278.2                   & \ldots                   \\
B004 & J004537.37$-$731236.50 & 11.40574 & $-$73.21014 &  3743.92 & 2976.41 & 1783.03 & 1008                     &  573.7                   &  391.1                   \\
B005 & J004600.13$-$732324.54 & 11.50056 & $-$73.39015 &  1502.89 & 1180.57 &  715.75 & \ldots                   & \ldots                   & \ldots                   \\
B006 & J004610.79$-$732535.61 & 11.54496 & $-$73.42656 &  2684.89 & 2484.81 & 1688.72 & 1339                     &  996.1                   &  852.6                   \\
B007 & J004611.72$-$732325.54 & 11.54884 & $-$73.39043 &  6530.19 & 5655.79 & 3742.47 & 2295.56\tablenotemark{a} & \ldots                   & \ldots                   \\
B009 & J004655.83$-$730452.89 & 11.73266 & $-$73.08136 &  3625.15 & 2824.18 & 1596.46 &  768.6                   &  523.2                   &  261.5                   \\
B010 & J004656.08$-$731054.83 & 11.73367 & $-$73.18190 &  8190.79 & 6787.24 & 4409.19 & 2667                     & 1709                     & 1199                     \\
\enddata
\tablenotetext{a}{Original S$^3$MC data point.}
\end{deluxetable}

\clearpage

%%%%%%% TABLE 2 - phottab2.txt
\begin{deluxetable}{lcccccccccccc}
\tabletypesize{\tiny}
\tablewidth{0pt}
\tablecolumns{13}
\tablecaption{SMC Dusty O and B Stars: Spitzer Data\label{tab:phottab2}}
\tablehead{
\colhead{ID No.} &
\colhead{3.6~\micron} &
\colhead{$\sigma_{3.6\micron}$} &
\colhead{4.5~\micron} &
\colhead{$\sigma_{4.5\micron}$} &
\colhead{5.8~\micron} &
\colhead{$\sigma_{5.8\micron}$} &
\colhead{8.0~\micron} &
\colhead{$\sigma_{8.0\micron}$} &
\colhead{24~\micron\tablenotemark{a}} &
\colhead{$\sigma_{24\micron}$\tablenotemark{a}} &
\colhead{70~\micron\tablenotemark{a}} &
\colhead{$\sigma_{70\micron}$\tablenotemark{a}} \\
&
\colhead{($\mu$Jy)} &
\colhead{($\mu$Jy)} &
\colhead{($\mu$Jy)} &
\colhead{($\mu$Jy)} &
\colhead{($\mu$Jy)} &
\colhead{($\mu$Jy)} &
\colhead{($\mu$Jy)} &
\colhead{($\mu$Jy)} &
\colhead{($\mu$Jy)} &
\colhead{($\mu$Jy)} &
\colhead{($\mu$Jy)} &
\colhead{($\mu$Jy)} \\
}
\startdata 
B001 & \ldots                   & \ldots &   444.40\tablenotemark{b} &   2.93 & \ldots                    & \ldots  & \ldots                     & \ldots  &  198102.30                  &   624.29 & \ldots                       & \ldots   \\
B003 &   58.53                  &   2.28 &    44.12                  &   2.51 & \ldots                    & \ldots  & \ldots                     & \ldots  &     930.76                  &    26.62 & \ldots                       & \ldots   \\
B004 &  159.10                  &   4.02 &    99.99                  &   3.94 &    98.28\tablenotemark{b} &    8.64 & \ldots                     & \ldots  &    1684.34                  &    26.40 & \ldots                       & \ldots   \\
B005 &  189.40                  &   7.51 &   168.70                  &   6.12 &   382.80                  &   23.00 &    666.20                  &   33.63 &    4215.01                  &    42.52 & \ldots                       & \ldots   \\
B006 &  263.10                  &  14.05 &   238.36\tablenotemark{b} &   1.74 &   287.10                  &   17.45 &    659.70                  &   40.85 &   22090.02                  &   111.60 & \ldots                       & \ldots   \\
B007 &  575.10                  &   9.76 &   412.60                  &   9.30 &   323.90                  &   14.23 &    198.40                  &   12.74 &    1647.90                  &    29.82 & \ldots                       & \ldots   \\
B009 &  142.30                  &   6.08 &    96.72                  &   4.76 & \ldots                    & \ldots  &     92.00\tablenotemark{b} &    9.48 &    2500.48                  &    31.08 & \ldots                       & \ldots   \\
B010 &  661.70                  &  14.72 &   500.50                  &  11.84 &   478.50                  &   18.28 &    584.80                  &   31.21 &    7382.77                  &    55.72 & \ldots                       & \ldots   \\
\enddata 
\tablenotetext{a}{Includes correction factor from MIPS to IRAC scale.}
\tablenotetext{b}{Original S$^3$MC data point.}
\end{deluxetable} 

\clearpage

%%%%% TABLE 3
\begin{deluxetable}{lcrrrccl}
\tabletypesize{\scriptsize}
\tablewidth{0pt}
\tablecolumns{8}
\tablecaption{Main Sequence Stellar Properties and Spectral Classification Scheme \label{tab:class}}
\tablehead{
\colhead{SpT} &
\colhead{T$_{\rm eff}$} &
\colhead{Radius} &
\colhead{Mass} &
\colhead{L$_{\rm bol}$} &
\colhead{Slope} &
\colhead{BC} &
\colhead{Criteria}\\
&
\colhead{($10^3$ K)}&
\colhead{(R$_\odot$)}&
\colhead{(M$_\odot$)}&
\colhead{(10$^3$ L$_\odot$)}\\
}
\startdata
O6     & 39 & 12.2  & 30.5 & 309.0 & $-$2.91 & $-$3.72 & \ion{He}{2} 4686, 4541 \AA, \ion{He}{1}+\ion{He}{2} 4026 $\sim$ \ion{He}{2} 4200 \AA \\
O7     & 37 & 11.3  & 27.3 & 213.3 & $-$2.91 & $-$3.54 & \ion{He}{2} 4686, 4200 \AA, \ion{He}{2} 4541 \AA~$\sim$ \ion{He}{1} 4471 \AA         \\
O8.5   & 34 &  9.9  & 22.8 & 118.4 & $-$2.91 & $-$3.27 & \ion{He}{2} 4686, 4200 \AA, \ion{He}{2} 4541 \AA~$\sim$ \ion{He}{1} 4387 \AA         \\
O9     & 33 &  9.5  & 21.5 & 96.3  & $-$2.91 & $-$3.18 & \ion{He}{2} 4686, \ion{He}{2} 4200 \AA~$\sim$ \ion{He}{1} 4143 \AA                   \\
B0     & 30 &  8.3  & 17.8 & 50.3  & $-$2.89 & $-$2.98 & \ion{He}{2} 4686, 4541 \AA, weak \ion{He}{2} 4200 \AA                                \\
B0.5   & 27 &  7.2  & 14.5 & 24.8  & $-$2.86 & $-$2.60 & Weak \ion{He}{2} 4686 \AA, no \ion{He}{2} 4200, 4541 \AA                             \\
B1/1.5 & 25 &  6.5  & 12.5 & 15.0  & $-$2.84 & $-$2.41 & \ion{Si}{4} 4088, 4116 \AA                                                           \\
B2/B3  & 22 &  5.6  & 10.0 & 6.6   & $-$2.79 & $-$2.11 & \ion{Si}{3} 4553 \AA                                                                 \\
B5/B8  & 15 &  3.8  & 5.4  & 0.6   & $-$2.67 & $-$1.24 & \ion{Si}{2} 4128/4132 \AA, no \ion{Si}{3} 4553 \AA                                   \\
B9     & 10 &  2.8  & 3.2  & 0.1   & $-$2.52 & $-$0.30 & \ion{Si}{2} 4128/4132 \AA, \ion{Mg}{2} 4481 \AA~$>$ \ion{He}{1} 4471 \AA             \\
\enddata
\end{deluxetable}

\clearpage

%%%%% TABLE 4 - excess1.txt + cross-identifications and additional info
\begin{deluxetable}{lccclcl}
\tabletypesize{\scriptsize}
\tablewidth{0pt}
\tablecolumns{7}
\tablecaption{SMC Dusty O and B Stars: Spectral Type, Radial Velocity and
Identification \label{tab:spec}}
\tablehead{
\multicolumn{2}{c}{This Work} &
\multicolumn{3}{c}{Literature} & & \\
\cline{3-5}
\colhead{No.} & \colhead{SpT\tablenotemark{a}} &
\colhead{SpT} & \colhead{v$_{rad}$} & \colhead{source\tablenotemark{b}} &
\colhead{v$_{rad}$ \ion{H}{1}} & \colhead{Identification} \\
 & & & \colhead{(km s$^{-1}$)} & & \colhead{(km s$^{-1}$)} & \\
}
\startdata
B001 &  Em    &                   &                        &             & & Planetary Nebula \\
B003 & B0--B2 &                   &                        &             & & \\
B004 &  O9    &                   &                        &             & & \\
B005 & \ldots &                   &                        &             & & Eclipsing binary \\
B006 & \ldots &                   &                        &             & & \\
B007 &  OBe   &                   &                        &             & & \\
B009 & B0--B2 &                   &                        &             & & \\
B010 &  OBe   &                   &                        &             & & Emission-line star \\
\enddata 
\tablenotetext{a}{"Em" denotes objects with forbidden emission lines. "OBe"
denotes Be stars. "OBe?" denotes stars with unusually weak \ha\ absorption. No
spectral type information indicates that the spectra were difficult to
extract.}
\tablenotetext{b}{
A09  = \citet{antoniou09};
C11  = \citet{coe11};
E06  = \citet{evans06};
EH08 = \citet{evans08};
H08  = \citet{hunter08};
M07  = \citet{martayan07};
S    = Simbad.}
\end{deluxetable} 

\clearpage

%%%%% TABLE 5 - excess1.txt
\begin{deluxetable}{lccc@{\extracolsep{10pt}}cc@{\extracolsep{10pt}}cc@{\extracolsep{10pt}}c}
% 24-Feb-2011 15:24:30
% using spec. type slope fit to J (or I if no J)
\tabletypesize{\tiny}
\tablewidth{0pt}
\tablecolumns{8}
\tablecaption{SMC Dusty O and B Stars: 4.5, 8, and 24~\micron\ Excesses
\label{tab:excess}}
\tablehead{
&
\multicolumn{2}{c}{4.5~\micron} &
\multicolumn{2}{c}{8.0~\micron} &
\multicolumn{2}{c}{24~\micron} &
 \\
\cline{2-3} \cline{4-5} \cline{6-7}
\colhead{No.} &
\colhead{Excess} &
\colhead{Ratio\tablenotemark{a}} &
\colhead{Excess} &
\colhead{Ratio\tablenotemark{a}} &
\colhead{Excess} &
\colhead{Ratio\tablenotemark{a}} &
\colhead{K--[24]} \\
&  
\colhead{($\mu$Jy)} & 
& 
\colhead{($\mu$Jy)} & 
& 
\colhead{($\mu$Jy)} & 
 & \\ 
}
\startdata
B001                  &   340$\pm$13  &  3.27$\pm$0.53 & \ldots          & \ldots         &  198098$\pm$624  & 41436$\pm$5181 & 10.86  \\
B003                  &     7$\pm$5   &  0.19$\pm$0.15 & \ldots          & \ldots         &     929$\pm$27   &   547$\pm$64   & \ldots \\
B004                  &    12$\pm$5   &  0.14$\pm$0.06 & \ldots          & \ldots         &    1681$\pm$26   &   467$\pm$19   &  6.49  \\
B005                  &   137$\pm$7   &  4.40$\pm$0.61 &    655$\pm$34   &  60.5$\pm$7.3  &    4214$\pm$43   &  2935$\pm$318  & \ldots \\
B006                  &   110$\pm$13  &  0.86$\pm$0.19 &    615$\pm$41   &  13.9$\pm$1.8  &   22084$\pm$112  &  3756$\pm$387  &  8.44  \\
B007                  &   193$\pm$26  &  0.88$\pm$0.21 &    122$\pm$15   &   1.6$\pm$0.3  &    1638$\pm$30   &   162$\pm$18   & \ldots \\
B009                  &    23$\pm$9   &  0.32$\pm$0.16 &     67$\pm$10   &   2.6$\pm$0.5  &    2497$\pm$31   &   740$\pm$80   &  7.36  \\
B010                  &   246$\pm$29  &  0.96$\pm$0.21 &    496$\pm$33   &   5.6$\pm$0.8  &    7371$\pm$56   &   629$\pm$65   &  6.88  \\
\enddata 
\tablenotetext{a}{Ratio is measured excess divided by expected photosphere.}
\tablenotetext{b}{Detected at 70~\micron.}
\end{deluxetable} 

\clearpage

%%%%% TABLE 6 - shortourdust1.tex
\begin{deluxetable}{lcccccccc}
% 07-Mar-2011 19:30:27
% using spec. type slope fit to J (or I if no J)
\tabletypesize{\tiny}
\tablewidth{0pt}
\tablecolumns{9}
\tablecaption{Dust Parameters for SMC Stars with Secure Spectral Types and Measurable 8 Micron Excess
\label{tab:ourdust}}
\tablehead{
\colhead{Object ID} &
\colhead{SpT} &
\colhead{T} &
\colhead{Mass} &
\colhead{F$_{IR}$/F$_{Bol}$} &
\colhead{$R_{\rm BB}$ \tablenotemark{a}} &
\colhead{$R_{\rm MBB}$ \tablenotemark{a}} &
\colhead{$a_{min}$} &
\colhead{$M_{\rm bo}$ \tablenotemark{b}} \\ 
&
&
\colhead{(K)} &
\colhead{(M$_\earth$)} &
\colhead{($\times 10^{-3}$)}&
\colhead{(au)} &
\colhead{($\times 10^4$ au)} &
\colhead{(mm)} &
\colhead{(M$_\earth$)} \\ 
}
\startdata
B021 &   B0  & 135$^{+4}_{-4}$ & 0.17$^{+0.05}_{-0.04}$ & 0.42$^{+0.31}_{-0.17}$ & 954$^{+124}_{-256}$ & 2.3 & 0.5 & 13 \\ 
B022 &   B0  & 132$^{+1}_{-1}$ & 1.08$^{+0.23}_{-0.21}$ & 2.30$^{+2.01}_{-0.97}$ & 994$^{+89}_{-261}$ & 2.5 & 0.5 & 80 \\ 
B023 &   O7  & 127$^{+0}_{-1}$ & 8.25$^{+1.67}_{-1.57}$ & 3.28$^{+0.89}_{-0.98}$ & 2211$^{+192}_{-168}$ & 5.8 & 1.3 & 1689 \\ 
B029 &   B0  & 118$^{+3}_{-3}$ & 1.82$^{+0.49}_{-0.37}$ & 1.95$^{+1.38}_{-0.80}$ & 1248$^{+130}_{-353}$ & 3.5 & 0.5 & 134 \\ 
B042 &   B0  & 136$^{+1}_{-1}$ & 1.12$^{+0.23}_{-0.21}$ & 2.88$^{+2.34}_{-1.16}$ & 933$^{+78}_{-254}$ & 2.3 & 0.5 & 82 \\ 
B043 &   O9  & 125$^{+3}_{-4}$ & 0.71$^{+0.19}_{-0.15}$ & 0.57$^{+0.30}_{-0.17}$ & 1526$^{+128}_{-240}$ & 4.0 & 0.8 & 83 \\ 
B047 &   O7  & 129$^{+2}_{-2}$ & 0.70$^{+0.16}_{-0.13}$ & 0.31$^{+0.08}_{-0.09}$ & 2144$^{+195}_{-183}$ & 5.5 & 1.3 & 144 \\ 
B048 &   B2  & 128$^{+3}_{-4}$ & 0.58$^{+0.14}_{-0.12}$ & 7.94$^{+3.35}_{-4.01}$ & 380$^{+92}_{-64}$ & 1.0 & 0.1 & 10 \\
\enddata 
\tablenotetext{a}{$R_{\rm BB}$ is the distance to the star assuming a blackbody emitter (Eq. \ref{bbd}),
$R_{\rm MBB}$ uses the modified blackbody emitter with $\lambda_0=1$~\micron\ (Eq. \ref{mbd}).}
\tablenotetext{b}{The mass of dust grains of size $a_{min}$, calculated
from F$_{IR}$/F$_{Bol}$ using Eq. \ref{Mbo}.}
\tablenotetext{c}{Has a 70~\micron\ detection.} 
\end{deluxetable} 

\clearpage

%%%%% TABLE 7 - shortgaudust1.tex
\begin{deluxetable}{lcccccccc}
% 01-May-2013 11:34:15 
% using spec. type slope fit to J (or I if no J) 
\tabletypesize{\tiny}
\tablewidth{0pt}
\tablecolumns{9}
\tablecaption{Gaustad and van Buren Dust Parameters \label{tab:gaudust}}
\tablehead{
\colhead{Object ID} &
\colhead{SpT} &
\colhead{T$_{12/24}$} & 
\colhead{Mass} &
\colhead{F$_{IR}$/F$_{Bol}$} &
\colhead{$R_{\rm BB}$ \tablenotemark{a}} &
\colhead{$R_{\rm MBB}$ \tablenotemark{a}} &
\colhead{$a_{min}$} &
\colhead{$M_{\rm bo}$ \tablenotemark{b}} \\
&
&
\colhead{(K)} & 
\colhead{(M$_{\earth}$)} & 
\colhead{($\times 10^{-3}$)} & 
\colhead{(AU)} & 
\colhead{($\times 10^4$ au)} & 
\colhead{(mm)} & 
\colhead{(M$_{\earth)}$}
}
\startdata
HR  533 &  B1  & 126 & 0.20 & 1.06 & 593 & 1.4 & 0.2 & 16 \\ 
HR 1191 &  B1  & 139 & 0.06 & 0.58 & 488 & 1.3 & 0.2 & 6 \\ 
HR 1763 &  B1  & 120 & 5.68 & 23.17 & 653 & 2.0 & 0.2 & 426 \\ 
HR 1798 &  B2  & 124 & 0.07 & 0.83 & 408 & 1.0 & 0.1 & 3 \\ 
HR 1855 &  B0  & 107 & 0.16 & 0.10 & 1522 & 4.6 & 0.5 & 23 \\ 
HR 1868 \tablenotemark{c} &  B1  & 124 & 0.48 & 2.38 & 614 & 1.8 & 0.2 & 39 \\ 
HR 1931 &  O9  & 103 & 2.39 & 0.58 & 2278 & 7.1 & 0.8 & 487 \\ 
HR 2058 &  B1  & 108 & 0.18 & 0.39 & 814 & 2.2 & 0.2 & 11 \\
\enddata 
\tablenotetext{a}{$R_{\rm BB}$ is the distance to the star assuming a blackbody emitter (Eq. \ref{bbd}),
$R_{\rm MBB}$ uses the modified blackbody emitter with $\lambda_0=1$~\micron\ (Eq. \ref{mbd}).}
\tablenotetext{b}{The mass of dust grains of size $a_{min}$, calculated
from F$_{IR}$/F$_{Bol}$ using Eq. \ref{Mbo}.}
\tablenotetext{c}{Known reflection nebula.} 
\end{deluxetable} 


\begin{thebibliography}{}
\bibitem[Adams et al.(2013)]{adams} Adams, J.~J., Simon, J.~D., Bolatto,
A.~D., Sloan, G.~C., Sandstrom, K.~M., Schmiedeke, A., van Loon, J.~Th.,
Oliveira, J.~M., \& Keller, L.~D.\ 2013, \apj submitted (Paper II)
\bibitem[Allen(1973)]{allen} Allen, C.~W.\ 1973, London: University of London,
Athlone Press
%, |c1973, 3rd ed.
% newer version, ed. by Cox ISBN: 978-0-387-95189-8
\bibitem[Alexander et al.(2006)]{alexander} Alexander, R.~D., 
Clarke, C.~J., \& Pringle, J.~E.\ 2006, \mnras, 369, 229 
\bibitem[Aller et al.(1982)]{kaler} Aller, L.~H., et al.\ 1982,
Landolt-Bornstein: Numerical Data and Functional Relationships in Science and
Technology
% ISBN: 978-3-540-10976-1
% 1996 version: ISBN: 978-3-540-56080-7
\bibitem[Andr{\'e} et 
al.(2010)]{andre} Andr{\'e}, P., Men'shchikov, A., Bontemps, S., et al.\ 2010, \aap, 518, L102 
\bibitem[Antoniou et al.(2009)]{antoniou09} Antoniou, V., Hatzidimitriou, D.,
Zezas, A., \& Reig, P.\ 2009, \apj, 707, 1080
\bibitem[Arny(1977)]{arny77} Arny, T.\ 1977, \apj, 217, 83
% \bibitem[Aumann et al.(1984)]{aumann} Aumann, H.~H., et al.\ 1984, \apjl,
% 278, L23
\bibitem[Backman \& Paresce(1993)]{back} Backman, D.~E., \& Paresce, F.\ 1993,
Protostars and Planets III, 1253
% \bibitem[Bolatto et al.(2000)]{dust2} Bolatto, A.~D., Jackson, J.~M.,
% Wilson, C.~D., \& Moriarty-Schieven, G.\ 2000, \apj, 532, 909
\bibitem[Bolatto et al.(2007)]{bolatto} Bolatto, A.~D., et al.\ 2007, \apj,
655, 212
\bibitem[Bonanos et al.(2010)]{sagesmc} Bonanos, A.~Z., et al.\ 2010, \aj,
140, 416
\bibitem[Burns et al.(1979)]{burns} Burns, J.~A., Lamy, P.~L., \& Soter, S.\
1979, Icarus, 40, 1
\bibitem[Carpenter et al.(2009)]{carp} Carpenter, J.~M., Mamajek, E.~E.,
Hillenbrand, L.~A., \& Meyer, M.~R.\ 2009, \apj, 705, 1646
\bibitem[Castelli \& Kurucz(2003)]{castelli} Castelli, F., \& Kurucz, R.~L.\
2003, Modelling of Stellar Atmospheres, 210, 20P
\bibitem[Cesaroni et al.(2007)]{cesaroni} Cesaroni, R., Galli, D., Lodato, G., Walmsley, C.~M., \& Zhang, Q.\ 2007, Protostars and Planets V, 197 
\bibitem[Chen et al.(2006)]{chen06} Chen, C.~H., et al.\ 2006, \apjs, 166, 351
% \bibitem[Currie et al.(2008)]{currie} Currie, T., Plavchan, P., \& Kenyon,
% S.~J.\ 2008, \apj, 688, 597
\bibitem[Cieza et al.(2010)]{cieza} Cieza, L.~A., Schreiber, 
M.~R., Romero, G.~A., et al.\ 2010, \apj, 712, 925 
\bibitem[Coe et al.(2011)]{coe11} Coe, M.~J., et al.\ 2011, \mnras, 414, 3281
\bibitem[Dawson et al.(2013)]{dawson} Dawson, J.~R., McClure-Griffiths, N.~M.,
Wong, T., Dickey, J.~M., Hughes, A., Fukui, Y., \& Kawamura, A.\ 2013, \apj,
763, 56
\bibitem[Dekker et al.(1986)]{emmi} Dekker, H., Delabre, B., \& Dodorico, S.\
1986, \procspie, 627, 339
\bibitem[Doyle et al.(2011)]{doyle} Doyle, L.~R., Carter, 
J.~A., Fabrycky, D.~C., et al.\ 2011, Science, 333, 1602 
\bibitem[Draine \& Lee(1984)]{dustk} Draine, B.~T., \& Lee, H.~M.\ 1984, \apj,
285, 89
\bibitem[Draine \& Li(2007)]{pahs} Draine, B.~T., \& Li, A.\ 2007, \apj, 657,
810
\bibitem[Evans et al.(2004)]{evans} Evans, C.~J., Howarth, I.~D., Irwin,
M.~J., Burnley, A.~W., \& Harries, T.~J.\ 2004, \mnras, 353, 601
\bibitem[Evans et al.(2006)]{evans06} Evans, C.~J., Lennon, D.~J., Smartt,
S.~J., \& Trundle, C.\ 2006, \aap, 456, 623
\bibitem[Evans \& Howarth(2008)]{evans08} Evans, C.~J., \& Howarth, I.~D.\
2008, \mnras, 386, 826
\bibitem[Fabregat \& Torrej{\'o}n(2000)]{fab} Fabregat, J., \& Torrej{\'o}n,
J.~M.\ 2000, \aap, 357, 451
\bibitem[Fazio et al.(2004)]{irac} Fazio, G.~G., et al.\ 2004, \apjs, 154, 10
\bibitem[Flower(1996)]{flower96} Flower, P.~J.\ 1996, \apj, 469, 355
% \bibitem[Flower(1977)]{flower} Flower, P.~J.\ 1977, \aap, 54, 31
\bibitem[Gaustad \& van Buren(1993)]{gaust} Gaustad, J.~E., \& van Buren, D.\
1993, \pasp, 105, 1127
\bibitem[Gordon et al.(2011)]{gordon11} Gordon, K.~D., et al.\ 2011, \aj, 142,
102
% \bibitem[Gorlova et al.(2006)]{gorl06} Gorlova, N., Rieke, G.~H., Muzerolle,
% J., Stauffer, J.~R., Siegler, N., Young, E.~T., \& Stansberry, J.~H.\ 2006,
% \apj, 649, 1028
\bibitem[Gvaramadze et al.(2011)]{gvaramadze} Gvaramadze, V.~V.,
Pflamm-Altenburg, J., \& Kroupa, P.\ 2011, \aap, 525A, 17
\bibitem[Herbig \& Simon(2001)]{hs01} Herbig, G.~H., \& Simon, T.\ 2001, \aj,
121, 3138
\bibitem[Hern{\'a}ndez et al.(2006)]{hern6} Hern{\'a}ndez, J., Brice{\~n}o,
C., Calvet, N., Hartmann, L., Muzerolle, J., \& Quintero, A.\ 2006, \apj, 652,
472
% \bibitem[Hern{\'a}ndez et al.(2009)]{hern9} Hern{\'a}ndez, J., Calvet, N.,
% Hartmann, L., Muzerolle, J., Gutermuth, R., \& Stauffer, J.\ 2009, \apj,
% 707, 705
% \bibitem[Hern{\'a}ndez et al.(2007)]{hern7} Hern{\'a}ndez, J., et al.\ 2007,
% \apj, 662, 1067
\bibitem[Hildebrand(1983)]{hildebrand83} Hildebrand, R.~H.\ 1983, 
\qjras, 24, 267 
\bibitem[Hillenbrand et al.(1993)]{hill2} Hillenbrand, L.~A., Massey, P.,
Strom, S.~E., \& Merrill, K.~M.\ 1993, \aj, 106, 1906
\bibitem[Hillenbrand et al.(1992)]{hill} Hillenbrand, L.~A., Strom, S.~E.,
Vrba, F.~J., \& Keene, J.\ 1992, \apj, 397, 613
\bibitem[Hunter et al.(2008)]{hunter08} Hunter, I., Lennon, D.~J., Dufton,
P.~L., Trundle, C., Sim\'on-D\'{\i}az, S., Smartt, S.~J., Ryans, R.~S.~I., \&
Evans, C.~J.\ 2008, \aap, 479, 541
\bibitem[Ita et al.(2010)]{akari} Ita, Y., et al.\ 2010, \pasj, 62, 273
% reference for blowout grain mass, from Morales et al
% \bibitem[Jura et al.(1995)]{1995ApJ...445..451J} Jura, M., Ghez, A.~M.,
% White, R.~J., McCarthy, D.~W., Smith, R.~C., \& Martin, P.~G.\ 1995, \apj,
% 445, 451
% \bibitem[Jura et al.(2004)]{jura} Jura, M., et al.\ 2004, \apjs, 154, 453
\bibitem[Kalas et al.(2002)]{neb} Kalas, P., Graham, J.~R., Beckwith,
S.~V.~W., Jewitt, D.~C., \& Lloyd, J.~P.\ 2002, \apj, 567, 999
\bibitem[Kaper et al.(1997)]{kaper} Kaper, L., van Loon, J.~Th., Augusteijn,
T., Goudfrooij, P., Patat, F., Waters, L.~B.~F.~M., \& Zijlstra, A.~A.\ 1997,
\apjl, 475, L37
\bibitem[Keller \& Wood(2006)]{kw06} Keller, S.~C., \& Wood, P.~R.\ 2006,
\apj, 642, 834
\bibitem[Krivov(2010)]{krivov} Krivov, A.~V.\ 2010, Research in Astronomy and
Astrophysics, 10, 383
% \bibitem[Li(2005)]{2005AIPC..761..123L} Li, A.\ 2005, The Spectral Energy 
% Distributions of Gas-Rich Galaxies: Confronting Models with Data, 761, 123
% "On the absorption and emission properties of interstellar grains"
\bibitem[Leroy et al.(2007)]{dusttogas} Leroy, A., Bolatto, A.,
Stanimirovi\'c, S., Mizuno, N., Israel, F., \& Bot, C.\ 2007, \apj, 658, 1027
\bibitem[Martayan et al.(2007)]{martayan07} Martayan, C., Fr\'emat, Y.,
Hubert, A.-M., Floquet, M., Zorec, J., \& Neiner, C.\ 2007, \aap, 462, 683
\bibitem[Mart{\'{\i}}nez-Galarza et al.(2009)]{mart} Mart{\'{\i}}nez-Galarza,
J.~R., Kamp, I., Su, K.~Y.~L., G{\'a}sp{\'a}r, A., Rieke, G., \& Mamajek,
E.~E.\ 2009, \apj, 694, 165
\bibitem[Martins et al.(2005)]{martins} Martins, F., Schaerer, D., \& Hillier,
D.~J.\ 2005, \aap, 436, 1049
\bibitem[Massey et al.(2005)]{massey} Massey, P., Puls, J., Pauldrach,
A.~W.~A., Bresolin, F., Kudritzki, R.~P., \& Snow, T.\ 2005, \apj, 627, 477
\bibitem[McSwain et al.(2009)]{mcswain} McSwain, M.~V., Huang, W., \& Gies,
D.~R.\ 2009, \apj, 700, 1216
% \bibitem[Meyer et al.(2007)]{meyer} Meyer, M.~R., Backman, D.~E.,
% Weinberger, A.~J., \& Wyatt, M.~C.\ 2007, Protostars and Planets V, 573
\bibitem[Meyssonnier \& Azzopardi(1993)]{meys} Meyssonnier, N., \& Azzopardi,
M.\ 1993, \aaps, 102, 451
% hd 4881 is a be star, also some interesting stuff on transition objects, but
% need an HR diagram
% \bibitem[Miroshnichenko et al.(1999)]{1999MNRAS.302..612M} Miroshnichenko, 
% A.~S., Mulliss, C.~L., Bjorkman, K.~S., Morrison, N.~D., Kuratov, K.~S., \&
% Wisniewski, J.~P.\ 1999, \mnras, 302, 612
\bibitem[Miroshnichenko \& Bjorkman(2000)]{mirbjor} Miroshnichenko, A.~S., \&
Bjorkman, K.~S.\ 2000, IAU Colloq.~175: The Be Phenomenon in Early-Type Stars,
214, 484
% \bibitem[Miroshnichenko et al.(2003)]{mir} Miroshnichenko, A.~S., Levato,
% H., Bjorkman, K.~S., \& Grosso, M.\ 2003, \aap, 406, 673
% does the blowout mass calculation for some A/late B stars
% \bibitem[Morales et al.(2009)]{2009ApJ...699.1067M} Morales, F.~Y., et al.\ 
% 2009, \apj, 699, 1067
\bibitem[Nieva(2013)]{nieva} Nieva, M.~F.\ 2013, \aap, 550, A26
\bibitem[Oliveira et al.(2013)]{oliveira} Oliveira, J.~M., et al.\ 2013,
\mnras, 428, 3001
\bibitem[Orosz et al.(2012a)]{orosz1} Orosz, J.~A., Welsh, 
W.~F., Carter, J.~A., et al.\ 2012, \apj, 758, 87 
\bibitem[Orosz et al.(2012b)]{orosz2} Orosz, J.~A., Welsh, 
W.~F., Carter, J.~A., et al.\ 2012, Science, 337, 1511 
\bibitem[Ostlie \& Carroll(1996)]{ostlie} Ostlie, D.~A., \& Carroll, B.~W.\
1996, An Introduction to Modern Stellar Astrophysics
\bibitem[Pagel et al.(1978)]{pagel78} Pagel, B.~E.~J., Edmunds, M.~G.,
Fosbury, R.~A.~E., \& Webster, B.~L.\ 1978, \mnras, 184, 569
\bibitem[Planck Collaboration et 
al.(2011)]{planck24} Planck Collaboration, Abergel, A., Ade, P.~A.~R., et al.\ 2011, \aap, 536, A24 
\bibitem[Porter \& Rivinius(2003)]{port} Porter, J.~M., \& Rivinius, T.\ 2003,
\pasp, 115, 1153
\bibitem[Rieke \& Lebofsky(1985)]{extum} Rieke, G.~H., \& Lebofsky, M.~J.\
1985, \apj, 288, 618
% 1-13 um extinction
\bibitem[Rieke et al.(2004)]{mips} Rieke, G.~H., et al.~2004, \apjs, 154, 25
% \bibitem[Rieke et al.(2005)]{rieke5} Rieke, G.~H., et al.\ 2005, \apj, 620,
% 1010
% \bibitem[Saffe et al.(2008)]{saffe} Saffe, C., G{\'o}mez, M., Pintado, O.,
% \& Gonz{\'a}lez, E.\ 2008, \aap, 490, 297
\bibitem[Sana et 
al.(2013)]{sana13} Sana, H., de Koter, A., de Mink, S.~E., et al.\ 2013, \aap, 550, A107 
\bibitem[Sana et al.(2012)]{sana12} Sana, H., de Mink, S.~E., 
de Koter, A., et al.\ 2012, Science, 337, 444 
\bibitem[Sandstrom et al.(2010)]{smcpah} Sandstrom, K.~M., Bolatto, A.~D.,
Draine, B.~T., Bot, C., \& Stanimirovi\'c, S.\ 2010, \apj, 715, 701
\bibitem[Schlegel et al.(1998)]{schlegel} Schlegel, D.~J., Finkbeiner, D.~P.,
\& Davis, M.\ 1998, \apj, 500, 525
% \bibitem[Sellgren et al.(1996)]{33um} Sellgren, K., Werner, M.~W., \&
% Allamandola, L.~J.\ 1996, \apjs, 102, 369
% \bibitem[Siegler et al.(2007)]{sieg} Siegler, N., Muzerolle, J., Young,
% E.~T., Rieke, G.~H., Mamajek, E.~E., Trilling, D.~E., Gorlova, N., \& Su,
% K.~Y.~L.\ 2007, \apj, 654, 580
\bibitem[Schwamb et al.(2012)]{schwamb} Schwamb, M.~E., Orosz, 
J.~A., Carter, J.~A., et al.\ 2012, arXiv:1210.3612 
\bibitem[Silaj et al.(2010)]{silaj} Silaj, J., Jones, C.~E., Tycner, C.,
Sigut, T.~A.~A., \& Smith, A.~D.\ 2010, \apjs, 187, 228
\bibitem[Skrutskie et al.(2006)]{mass} Skrutskie, M.~F., et al.\ 2006, AJ,
131, 1163
\bibitem[Sloan et al.(2004)]{sloan04} Sloan, G.~C., et al.\ 2004, \apjl, 614,
L77
\bibitem[Smith \& Wyatt(2010)]{smwy} Smith, R., \& Wyatt, M.~C.\ 2010, \aap,
515A, 95
\bibitem[Sota et al.(2011)]{sota} Sota, A., Ma\'{\i}z-Apell\'aniz, J.,
Walborn, N.~R., Alfaro, E.~J., Barb\'a, R.~H., Morrell, N.~I., Gamen, R.~C.,
Arias, \& J.~I.\ 2011, \apjs, 193, 24
\bibitem[Stanimirovi\'c et al.(1999)]{stanimirovic99} Stanimirovi\'c, S.,
Staveley-Smith, L., Dickey, J.~M., Sault, R.~J., \& Snowden, S.~L.\ 1999,
\mnras, 302, 417
\bibitem[Stanimirovi\'c et al.(2004)]{stanimirovic04} Stanimirovi\'c, S.,
Staveley-Smith, L., \& Jones, P.~A.\ 2004, \apj, 604, 176
% \bibitem[Stanimirovi\'c et al.(2005)]{dust1} Stanimirovi\'c, S., Bolatto,
% A.~D., Sandstrom, K., Leroy, A.~K., Simon, J.~D., Gaensler, B.~M., Shah,
% R.~Y., \& Jackson, J.~M.\ 2005, \apjl, 632, L103
% \bibitem[Stanimirovi\'c et al.(2000)]{stan2000} Stanimirovi\'c, S.,
% Staveley-Smith, L., van der Hulst, J.~M., Bontekoe, T.~R., Kester, D.~J.~M.,
% \& Jones, P.~A.\ 2000, \mnras, 315, 791 replaced with Leroy et al, 2007
% {dusttogas}
% 51 Oph - a beta pic analog with halpha emission
% \bibitem[Stark et al.(2009)]{2009ApJ...703.1188S} Stark, C.~C., et al.\
% 2009, \apj, 703, 1188
% \bibitem[Stauffer et al.(2005)]{stauff} Stauffer, J.~R., et al.\ 2005, \aj,
% 130, 1834
\bibitem[Staveley-Smith et al.(1997)]{lister} Staveley-Smith, L., Sault,
R.~J., Hatzidimitriou, D., Kesteven, M.~J., \& McConnell, D.\ 1997, \mnras,
289, 225
% \bibitem[Su et al.(2005)]{surprise} Su, K.~Y.~L., et al.\ 2005, \apj, 628,
% 487
\bibitem[Su et al.(2006)]{su06} Su, K.~Y.~L., et al.\ 2006, \apj, 653, 675
\bibitem[Takeuchi \& Artymowicz(2001)]{tak} Takeuchi, T., \& Artymowicz, P.\
2001, \apj, 557, 990
\bibitem[Terebey \& Fich(1986)]{tere} Terebey, S., \& Fich, M.\ 1986, \apjl,
309, L73
\bibitem[Testor(2001)]{testor} Testor, G.\ 2001, \aap, 372, 667
% \bibitem[The et al.(1994)]{the} The, P.~S., de Winter, D., \& Perez, M.~R.\
% 1994, \aaps, 104, 315
\bibitem[Touhami et al.(2010)]{beseds} Touhami, Y., et al.\ 2010, \pasp, 122,
379
\bibitem[Udalski et al.(1998)]{ogle} Udalski, A., Szyma\'nski, M., Kubiak, M.,
Pietrzy\'nski, G., Wo\'zniak, P., \& $\dot{\rm Z}$ebru\'n, K.\ 1998, Acta
Astronomica, 48, 147
% the GLIMPSE paper with the flow chart (fig 18) that distinguishes transition
% disks, classical Be stars, HAeBe stars, and debris disks depending on
% emission/ lack of emission features
% \bibitem[Uzpen et al.(2008)]{2008ApJ...685.1157U} Uzpen, B., Kobulnicky,
% H.~A., Semler, D.~R., Bensby, T., \& Thom, C.\ 2008, \apj, 685, 1157
\bibitem[van Buren \& McCray(1988)]{vanmc} van Buren, D., \& McCray, R.\ 1988,
\apjl, 329, L93
\bibitem[van Loon et al.(2010)]{vanloon10} van Loon, J.~Th., Oliveira, J.~M.,
Gordon, K.~D., Sloan, G.~C., \& Engelbracht, C.~W.\ 2010, \aj, 139, 1553
%\bibitem[van Loon et al.(2013)]{vanloon13} van Loon, J.~Th., et al.\ 2013,
%\aap, in press (arXiv:1210.4447)
% \bibitem[Walborn \& Fitzpatrick(1990)]{walfitz} Walborn, N.~R., \&
% Fitzpatrick, E.~L.\ 1990, \pasp, 102, 379
\bibitem[Waters \& Marlborough(1994)]{waters0} Waters, L.~B.~F.~M., \&
Marlborough, J.~M.\ 1994, IAU Symposium 162: Pulsation; rotation; and mass
loss in early-type stars (Kluwer), 399
\bibitem[Waters \& Waelkens(1998)]{waters} Waters, L.~B.~F.~M., \& Waelkens,
C.\ 1998, \araa, 36, 233
\bibitem[Welsh et al.(2012)]{welsh} Welsh, W.~F., Orosz, 
J.~A., Carter, J.~A., et al.\ 2012, \nat, 481, 475 
\bibitem[Werner et al.(2004)]{spitzer} Werner, M.~W., et al.\ 2004, \apjs,
154, 1
\bibitem[White \& Bally(1993)]{wb93} White, R.~E., \& Bally, J.\ 1993, \apj,
409, 234
\bibitem[Zaritsky et al.(2002)]{zar} Zaritsky, D., Harris, J., Thompson,
I.~B., Grebel, E.~K., \& Massey, P.\ 2002, \aj, 123, 855
\end{thebibliography}
\end{document}